\documentclass[preprint,10pt]{elsarticle} 
\usepackage{amssymb}
\usepackage{amsmath}
\usepackage{color}
\usepackage{parskip}
\usepackage{comment}

\biboptions{sort}

\usepackage[margin=1in]{geometry}
\usepackage{setspace}

\newcommand{\Red}{\color{red}}
\newcommand{\Blue}{\color{blue}}
\newcommand{\bOmega}{\boldsymbol{\Omega}}
\newcommand{\bD}{\boldsymbol{D}}
\newcommand{\lapl}{\triangle}

\journal{Journal of Computational Physics}

\begin{document}

\onehalfspacing

\title{A fast, high-order solver for the Grad-Shafranov equation}

\author[cims]{Andras Pataki\fnref{fn1}}
\ead{apataki@apataki.net}

\author[cims]{Antoine J. Cerfon\fnref{fn2}}
\ead{cerfon@cims.nyu.edu}

\author[mit]{Jeffrey P. Freidberg\fnref{fn2}}
\ead{jpfreid@mit.edu}

\author[cims]{Leslie Greengard\fnref{fn1}}
\ead{greengard@cims.nyu.edu}

\author[cims]{Michael O'Neil\fnref{fn1}}
\ead{oneil@cims.nyu.edu}

\address[cims]{Courant Institute of Mathematical Sciences,
New York University, New York, NY 10012}
\address[mit]{Plasma Science and Fusion Center, Massachusetts 
Institute of Technology, Cambridge, MA 02139}


\fntext[fn1]{Research supported in part by the National Science
Foundation under grant DMS06-02235, the U.S. Department of Energy
under contract DE-FG02-88ER-25053, and the Air Force Office of 
Scientific Research under NSSEFF Program Award FA9550-10-1-0180.}

\fntext[fn2]{Research supported in part by U.S. Department of Energy
under contract DE-FG02-91ER-54109.}

\begin{abstract}
We present a new fast
solver to calculate fixed-boundary plasma equilibria in toroidally axisymmetric geometries. By combining conformal mapping with Fourier and integral equation methods on the unit disk, we show that high-order accuracy can be achieved for the solution of the equilibrium equation and its first and second derivatives. Smooth arbitrary plasma cross-sections as well as arbitrary pressure and poloidal current profiles are used as initial data for the solver. 
Equilibria with large Shafranov shifts can be computed without difficulty. Spectral convergence is demonstrated by comparing the numerical solution with a known exact analytic solution. A fusion-relevant example of an equilibrium 
with a pressure pedestal is also presented.
\end{abstract}


\maketitle

\section{Introduction}

High performance numerical equilibrium solvers are crucial to the 
computational effort in plasma physics for magnetic fusion applications.
Numerically computed plasma equilibria are needed as an input 
to the algorithms responsible for calculating
the macroscopic stability and transport properties of plasma configurations.
Equilibrium solvers need to be fast, so that computational time can 
primarily be spent on the stability and transport calculations.
High accuracy is also required in order to minimize error propagation.

The equilibrium magnetic configuration in all toroidally axisymmetric magnetic confinement devices, including the tokamak and the spherical tokamak (ST), is determined by solving the Grad-Shafranov (GS)
equation \cite{Grad,Shafranov,Schluter} (also known as the Grad-Schl\"uter-Shafranov equation).
The Grad-Shafanov equation is a nonlinear elliptic partial differential equation, which, in general, can only be solved numerically. Because of its importance in computational plasma physics, a wide range of numerical methods have been developed \cite{Takeda_review}.
These schemes generally fall into two categories.
Eulerian or ``direct'' solvers use a prescribed mesh to calculate the 
unknown function 
\cite{Holmes_Euler,Huysmans_Euler,Lutjens_Euler,Jardin_Euler,Gourdain_Euler}, while Lagrangian or ``inverse'' solvers find the mapping of the plasma geometry in terms of magnetic coordinates 
\cite{DeLucia_inverse,Ling_inverse,Gruber_inverse,Turkington_inverse,Ludwig_inverse}.
The advantages and disadvantages of one formulation as compared to the other depend on the application of interest \cite{Zakharov}, on the plasma geometry, and on the type of inputs that plasma stability and transport codes require.
For the Eulerian formulations, finite difference 
\cite{Holmes_Euler,Gourdain_Euler} and finite element 
\cite{Huysmans_Euler,Lutjens_Euler,Jardin_Euler,Lutjens_CHEASE2,Howell_NIMEQ}
methods are systematically favored.

In this article we present a new, high-order, fixed boundary, 
direct GS solver and demonstrate its effectiveness as part of 
an Eulerian solver.
High-order accuracy is achieved through the combination of three key elements:
a rescaling of the unknown which reformulates the GS equation as a
nonlinear Poisson problem, a spectrally-accurate numerical method to 
compute the conformal map from the plasma cross-section to the unit disk,
and finally, a fast, high-order Poisson solver on the unit disk. 

Conformal mapping techniques in the context of numerical equilibrium solvers in plasma physics were first considered by Goedbloed \cite{Goedbloed_conform_1,Goedbloed_conform_2,Goedbloed_textbook} as a convenient way to decouple the numerical issues associated with the plasma geometry from the rest of the problem.
We show that for smooth-boundary plasma cross-sections, spectrally-accurate conformal maps to the unit disk can be efficiently computed.
It is important to note that the map needs to be computed only once for a particular geometry, and can be used to iteratively solve the nonlinear equation or to compute equilibria for several plasma profiles.
By combining this mapping with a Green's function-based Poisson solver on the unit disk, we obtain high-order accuracy for the solution as well as its derivatives. This is one of the main motivations for this work, since the stability and transport properties of the plasma are very sensitive to quantities such as the local magnetic shear and the magnetic field curvature which depend on the second derivatives of the GS solution \cite{Freidberg_book}.

The structure of this article is as follows. In Section 2, we introduce the Grad-Shafranov equation and give the mathematical formulation of the fixed-boundary problem. In Section 3, we present a rescaling of the unknown 
which reformulates the GS equation as a nonlinear Poisson problem, 
and describe the iterative method used for its solution.
Section 4 contains the details of the numerical conformal mapping scheme
which maps the plasma domain to the unit disk. The conformal map is computed
by solving the Kerzman-Stein integral equation, and using its relation
to the Szeg\H o kernel.
The fast Poisson solver on the unit disk is presented in Section 5.
In Section 6, we illustrate the efficiency and accuracy of our new solver with a few specific examples, and in Section 7 we conclude with a discussion of the limitations of the solver, and its potential extensions.

\section{The Grad-Shafranov equation as a nonlinear eigenvalue problem}
\label{sec:GS_presentation}

\subsection{The Grad-Shafranov equation}
In toroidally axisymmetric systems,
in the usual cylindrical coordinate system $(R,\phi,Z)$,
the magnetic field can be expressed as
\begin{equation}
\vec{B}=\frac{g(\Psi)}{R}\vec e_{\phi}+\frac{1}{R}\nabla\Psi\times\vec e_{\phi},
\label{eq:basic_field}\end{equation}
where $\vec e_{\phi}$ is a unit vector in the toroidal direction.
By the assumption of axisymmetry, none of the physical functions of interest
depend on the angle $\phi$.
Here  $2\pi\Psi(R,Z)$ represents the poloidal magnetic flux,
and $2\pi g(\Psi)=-I_{p}(\Psi)$ is the net poloidal current flowing in the plasma and the toroidal field coils.
The flux function $\Psi$ satisfies the Grad-Shafranov equation
\begin{equation}
\lapl^*\Psi = R\frac{\partial}{\partial R}\left(\frac{1}{R}\frac{\partial\Psi}{\partial R}\right)+\frac{\partial^{2}\Psi}{\partial Z^{2}}=-\mu_{0}R^{2}\frac{dp}{d\Psi}-\frac{1}{2}\frac{dg^{2}}{d\Psi} ,
\label{eq:GS_eq}
\end{equation}
where $p$ is the plasma pressure. Both $p$ and $g$ are application-specific
functions of $\Psi$, which, along with the boundary conditions, 
determine the equilibrium.
Indeed, once (\ref{eq:GS_eq}) is solved and $\Psi$ is known, 
$p$ and $g$ can be immediately evaluated, $\vec{B}$ can be computed from (\ref{eq:basic_field}), and the current density $\vec{J}$ in the plasma is 
given by (see \cite{Freidberg_book}, for example):
\begin{align}
	\vec{J}
        &=
        \frac{1}{\mu_0 R}\frac{dg}{d\Psi}{\nabla}\Psi\times\vec e_{\phi}
        -\frac{1}{\mu_0 R}\lapl^*\Psi\vec e_{\phi} \, .
\label{eq:basic_current}
\end{align}

\subsection{Boundary conditions}
Equation (\ref{eq:GS_eq}) is a second-order elliptic nonlinear partial differential equation.
Depending on the boundary conditions to be enforced, we distinguish
between two general classes of problems: (i) fixed-boundary problems in which
the plasma boundary is prescribed, with $\Psi = constant$
on the boundary and one solves for $\Psi$ inside the plasma, 
and (ii) free-boundary problems where
the current flowing in a set of external coils is given
and one has to find $\Psi$ such that the equilibrium is 
self-consistent with these currents.
In this paper, we consider only fixed-boundary problems.
That is, the shape of the plasma boundary is given 
and $\Psi=constant$ on the boundary.
Furthermore, since only derivatives of $\Psi$ represent physical quantities, we can choose $\Psi=0$ on the boundary without loss of generality.

Our motivation for focusing on the fixed-boundary problem is two-fold.
First, a large number of plasma stability and transport numerical codes take fixed boundary equilibria as their initial data.
In particular, parametric studies to understand and optimize plasma 
properties by varying a few parameters defining a given generic plasma boundary
are very common.
Second, many 
free-boundary GS solvers use iterative schemes which require a 
robust fixed-boundary solver in the iterative 
loop \cite{Takeda_review,Jardin_textbook}.

\subsection{The eigenvalue problem}
As explained in the previous section,
the generic form of the fixed-boundary GS equation can be written as
\begin{align}
  \begin{aligned}
	\lapl^* \Psi &= F(\Psi, R, Z)
        && \text{in}\;\bOmega
        \\
        \Psi &= 0
        && \text{on}\;\partial \bOmega,
  \end{aligned}
  \label{eq:GS_1}
\end{align}
with
\begin{align*}
        F(\Psi, R, Z) 
&= -\mu_{0}R^{2}\frac{dp}{d\Psi}-\frac{1}{2}\frac{dg^{2}}{d\Psi},
\end{align*}
where $\bOmega$ represents the interior of the plasma, 
and $\partial\bOmega$ the plasma boundary.
The nature of the previous problem depends on the behavior of the right hand
side as a function of $\Psi$; the user-specified plasma
profiles determine this. For many profiles, the right hand side
$F$ is of the form
\begin{equation}
F(\Psi,R,Z)=\tilde{F}(\Psi,R,Z) \, \Psi.
\end{equation}
In this case, the function $\Psi = 0$ is a trivial, 
but not physically relevant, solution to equation (\ref{eq:GS_1}).
It is well-known \cite{Takeda_review,Goedbloed_conform_2,Lodestro} that in this case the Grad-Shafranov equation is solved as an eigenvalue problem,
which is linear when $F$ is linear in $\Psi$, and nonlinear otherwise.
In particular, consider the scalings
\begin{align}
	\Psi &= |\Psi|_{m}\bar{\Psi},
        &
        \frac{dp}{d\Psi} &= \frac{\sigma}{|\Psi|_{m}}\frac{d\bar{p}(\bar{\Psi})}{d\bar{\Psi}} ,
        \label{eq:psi_p_normal}
        \\
        \frac{dg^{2}}{d\Psi} &= \frac{\sigma}{|\Psi|_{m}}\frac{d\bar{g}^{2}(\bar{\Psi})}{d\bar{\Psi}},
        &
        F(\Psi,R,Z) &= \frac{\sigma}{|\Psi|_{m}}\bar{F}(\bar{\Psi},R,Z) ,
        \label{eq:g_f_normal}
\end{align}
where $|\Psi|_{m} = \sup|\Psi|$
and $\sigma$ is a normalization factor for
the pressure and poloidal current profiles.
Note that $\bar{\Psi}$, as defined in (\ref{eq:psi_p_normal}), takes values in the interval $[-1, 1]$.
Inserting the normalizations in equations~(\ref{eq:psi_p_normal}) 
and~(\ref{eq:g_f_normal}) into~(\ref{eq:GS_1}), we find
\begin{equation}
\lapl^*\bar{\Psi}=\frac{\sigma}{|\Psi|_{m}^{2}}\bar{F}(\bar{\Psi},R,Z).
\label{eq:GS_2}
\end{equation}
Equation (\ref{eq:GS_2}) is in the desired form, highlighting a well-known property of the GS equation:
its scale-invariance under the transformation
\begin{equation}
\left(\Psi,\sigma,R,Z \right) \rightarrow \left( \lambda\Psi, 
\lambda^2 \sigma, R, Z \right).
\label{eq:invariance}
 \end{equation}
This scale invariance implies that by defining the ratio 
$\bar{\sigma}=\sigma/|\Psi|_{m}^{2}$, we can write the fixed-boundary 
GS equation in the following scale-independent form:
\begin{align}
  \begin{aligned}
    \lapl^{*}\bar{\Psi} &= \bar{\sigma}\bar{F}(\bar{\Psi},R,Z)
    && \mbox{in}\;\;\bOmega
    \\
    \bar{\Psi} &=0 
    && \mbox{on}\;\;\partial\bOmega.
  \end{aligned}
  \label{eq:GS_3}
\end{align}
This is now an eigenvalue problem to be solved for the 
eigenfunction $\bar{\Psi}$ with eigenvalue $\bar{\sigma}$.
When transforming back to the original variables that represent 
physical quantities of interest, there are two choices.
Either $|\Psi|_{m}$ is given (or alternatively $I_{\phi}$, 
the total toroidal current flowing in the plasma), and the magnitude 
of the pressure and current profiles is computed from the 
relation $\sigma=|\Psi|_{m}^{2}\bar{\sigma}$,
or the normalization $\sigma$ is given 
and the corresponding $|\Psi|_{m}$ or $I_{\phi}$ is determined.

On the other hand, the pressure profile 
or the poloidal current profile may be specified to include
terms which are linear in $\Psi$. Under this assumption, their derivatives 
can be written in the form
\begin{equation}
	-\mu_{0}\frac{dp(\Psi)}{d\Psi}=
        -\mu_{0}S(\Psi)\Psi+C, \qquad
        -\frac{dg^{2}}{d\Psi}=
        T(\Psi)\Psi+A,
	\label{eq:profiles_with_constant}
\end{equation}
with $A$ and $C$ constants such that $C + A \neq 0$.
This choice of $p$ and $g$ result in the right hand side $F$ of 
equation~(\ref{eq:GS_1}) having a constant term. This choice
corresponds to a discontinous toroidal current $J_{\phi}$ at the plasma edge,
and equation~(\ref{eq:GS_1}) does \emph{not} correspond to 
an eigenvalue problem.
While this is not a physically relevant choice,
it plays an important role in the benchmarking of numerical 
Grad-Shafranov solvers.
When the functions $S$ and $T$ are identically zero (corresponding to 
the so-called \emph{Solov'ev profiles} \cite{Solovev})
exact analytic solutions to the GS equation with fixed boundary conditions 
can be constructed \cite{Freidberg_book,Zheng,Cerfon},
which can then be used to test the accuracy of a numerical solver.
In Section \ref{sec:tests} we present convergence results
of our solver benchmarked against such analytic solutions.

In what follows, we describe a new numerical 
scheme which solves equation (\ref{eq:GS_1}) with high accuracy.
Both the eigenvalue and non-eigenvalue problems will be considered.
For the eigenvalue problem, we only treat the question of finding the 
smallest eigenvalue $\bar{\sigma}$ and its associated eigenfunction
because almost all magnetic fusion confinement problems of interest 
have a single extremum of $\Psi$ within the plasma region.
Our procedure could however be easily generalized to find other eigenvalues, 
for so-called \emph{doublet} or more ornate plasma configurations.

\section{The Grad-Shafranov equation as an iterative Poisson problem}
\label{sec:iteration}

\subsection{Transformation to a two-dimensional nonlinear Poisson problem}

We consider the generic form of the fixed-boundary Grad-Shafranov equation, given by equation (\ref{eq:GS_1}).
In all but the simplest cases, an iterative scheme is necessary
since either $F(\Psi,R,Z)$ is a nonlinear function of $\Psi$,
or a nonlinear eigenvalue problem is to be solved (or both).
As defined earlier, 
the differential operator $\lapl^*$ is a second-order elliptic operator,
similar to the Laplacian:
\begin{align*}
	\lapl^* \Psi
	&=
	\Psi_{RR} - \frac{1}{R} \Psi_R + \Psi_{ZZ}.
\end{align*}
If iteration is to be used, 
one might consider rewriting equation (\ref{eq:GS_1}) as
\begin{align}
  \begin{aligned}
	\Psi_{RR}+\Psi_{ZZ}
        &=
        F(\Psi,R,Z)+\frac{1}{R}\Psi_{R}
        \hspace{16pt}
        && \text{in}\;\bOmega
        \\
        \Psi &= 0
        && \text{on}\;\partial \bOmega,
  \end{aligned}
  \label{eq:bad_transfo}
\end{align}
and solve iteratively using a two-dimensional Cartesian Poisson solver 
at each step, holding
the right hand side fixed at the previous iterate.
However, this approach has one caveat.
It is empirically observed that fixed-point iteration converges 
faster when the right-hand side is slowly varying.
It would therefore be preferable to have a right-hand side 
that is only dependent on the unknown function itself, 
and not on its derivatives.
In order to obtain this desired form, we consider the following 
scaling of the unknown function $\Psi$:
\begin{equation}
	U(R,Z) = \frac{1}{\sqrt{R}} \Psi(R,Z).
	\label{eq:Laplace_transfo}
\end{equation}
This transformation is clearly singular at $R=0$.
Fortunately, this singularity does not have any consequence 
because in all applications of interest,
the physical domain excludes the $Z$-axis.
Inserting transformation~(\ref{eq:Laplace_transfo})
into~(\ref{eq:GS_1}), we readily find that $U$ satisfies
the following equation:
\begin{align}
  \begin{aligned}
	U_{RR}+U_{ZZ} &= \mathcal{F}(U,R,Z) \hspace{16pt}
        && \text{in}\;\bOmega
        \\
        U &= 0
        && \text{on}\;\partial \bOmega \, ,
  \end{aligned}
  \label{eq:GS_transfo}
\end{align}
where $\mathcal F$ is defined as
\begin{align*}
	\mathcal{F}(U, R, Z)
	&=
        \frac{1}{\sqrt{R}}F(\sqrt{R}U,R,Z)+\frac{3}{4R^{2}}U.
\end{align*}

Here, $\mathcal{F}$ is a function of $U$ only, and not of
its derivatives.
Equation~(\ref{eq:GS_transfo}) has the form of a 
two-dimensional nonlinear Poisson problem
(treating $R$ and $Z$ as Cartesian coordinates),
which we solve iteratively as described in the next section.
Once $U$ and all its derivatives have been computed,
all quantities of physical interest can be evaluated
to the same accuracy using the relations:
\begin{align*}
	\Psi_{R} & = \frac{1}{2\sqrt{R}}U+\sqrt{R}U_{R}
        \\
        \Psi_{Z} &= \sqrt{R}U_{Z}
        \\
        \Psi_{RR} &= -\frac{1}{4R^{3/2}}U+\frac{1}{\sqrt{R}}U_{R}+\sqrt{R}U_{RR}
        \\
        \Psi_{ZZ} &= \sqrt{R}U_{ZZ}.
\end{align*}

\subsection{Iterative solution for the Grad-Shafranov equation}

When the nonlinear Poisson problem defined by~(\ref{eq:GS_transfo}) 
does not correspond to an eigenvalue problem (i.e. the right
hand side contains a constant term),
we use a straight-forward fixed-point iteration scheme.
At each iteration, the right-hand side of (\ref{eq:GS_transfo}) 
is evaluated at the previous value of $U$,
and the Laplacian 
$\lapl = \partial^{2}/\partial R^{2}+\partial^{2}/\partial Z^{2}$
is inverted to calculate the updated value of $U$.
In other words, on the $l^{th}$ iteration, we solve
\begin{equation}
\lapl U_{l}=\mathcal{F}(U_{l-1},R,Z)
\end{equation}
for $U_{l}$. The iterative process is stopped when 
$||U_{l}-U_{l-1}||_\infty < \epsilon$ for a few consecutive iterations,
with $\epsilon$ small and chosen by the user.

The iterative scheme for the related eigenvalue problem is more subtle.
Equation~(\ref{eq:GS_3}) describes a \textit{nonlinear} eigenvalue problem
to be solved for the smallest eigenvalue $\bar{\sigma}$ and its 
associated eigenfunction
$\bar{\Psi}$, with $||\bar{\Psi}||_{\infty}=1$.
Here, the norm $||\cdot||_{\infty}$ has its usual meaning, i.e.
 $||\bar{\Psi}||_{\infty}=\sup(|\Psi|)$.
When solving a linear eigenvalue problem, 
the eigenvalue does not depend on the scaling of the eigenfunction.
However, this is generally not true for nonlinear problems.
Therefore, we use a modified version of the well-known inverse iteration method \cite{Trefethen}
that evolves the eigenvalue-eigenfunction pair simultaneously
instead of evolving the eigenfunction alone.
First, the previous iterate eigen-pair $(\bar{\Psi}_{l-1}, \bar{\sigma}_{l-1})$
is used to solve the PDE
\begin{align}
  \begin{aligned}
	\lapl^* \Psi_{l}
	&= \bar{\sigma}_{l-1} F(\bar{\Psi}_{l-1}, R, Z) \hspace{16pt}
        && \text{in}\;\bOmega
        \\
        \Psi_{l} &= 0
        && \text{on}\;\partial \bOmega \, .
  \end{aligned}
  \label{eq:eigenvalue_psi_1}
\end{align}
Then, the next eigen-pair iterate $(\bar{\Psi}_{l}, \bar{\sigma}_{l})$ 
is computed via correcting by the norm of $\Psi_l$:
\begin{equation}
  	\bar{\Psi}_{l} = \frac{\Psi_{l}}{||\Psi_{l}||_{\infty}}, \qquad
        \bar{\sigma}_{l} = \frac{\bar{\sigma}_{l-1}}{||\Psi_{l}||_{\infty}}
        \label{eq:eigenvalue_psi_2} 
\end{equation}
Once again, the iterative process can be terminated
when $||\bar{\Psi}_{l}-\bar{\Psi}_{l-1}||_\infty < \epsilon$
for a few consecutive iterations, with $\epsilon$ small and chosen 
by the user.

Applying the transformation (\ref{eq:Laplace_transfo})
to (\ref{eq:eigenvalue_psi_1}) and (\ref{eq:eigenvalue_psi_2}) gives the iterative scheme in terms of  $U$.
First we solve the Poisson problem for $\mathcal{U}_l$:
\begin{align}
  \begin{aligned}
	\lapl\mathcal{U}_{l}
        &=
        \frac{\bar{\sigma}_{l-1}}{\sqrt{R}}F(\sqrt{R}U_{l-1},R,Z)+\frac{3}{4R^{2}}U_{l-1}
        \hspace{12pt}
        &&  \text{in}\;\bOmega
        \\
        \mathcal{U}_{l}
        &=
        0
        && \text{on}\; \partial\bOmega \, .
  \end{aligned}
\label{eq:nonlinear_eigen}
\end{align}
Then, we compute the updated $U_l$ and $\bar{\sigma}_l$ as:
\begin{equation}
	U_{l}
        =
        \frac{\mathcal{U}_{l}}{||\sqrt{R}\,\mathcal{U}_{l}||_{\infty}},
        \qquad
        \bar{\sigma}_{l}
        =
        \frac{\bar{\sigma}_{l-1}}{||\sqrt{R}\,\mathcal{U}_{l}||_{\infty}}
        \label{eq:BC_nonlinear_eigen_3}
\end{equation}
This eigenvalue iteration scheme 
generally converges at a geometric rate, as shown in \cite{Pataki}.
Note that in this scheme, we calculate $||\sqrt{R}\,\mathcal{U}_{l}||_{\infty}$ 
by first finding the maximum on the computational grid,
and then using Newton's method to refine the maximum.
Without this latter step, the problem we solve would depend on the 
computational grid
and would converge
only linearly, and not spectrally approximate $||\sqrt{R}\,U_l||_\infty$
as the grid is refined.

In summary, we have shown that the Grad-Shafranov equation can be solved
using a sequence of iterations which require 
a single two-dimensional Poisson boundary value solve
in Cartesian coordinates at each step.
We now move on to developing a highly accurate Poisson solver
in $\bOmega$ with Dirichlet conditions $U=0$ on $\partial\bOmega$.
This will be accomplished in two steps.
First, we present a fast spectrally-accurate method for 
numerically computing the conformal map
from our domain $\bOmega$ to the unit disk.
Then, we develop a fast, high-order Poisson solver on the unit disk
using separation of variables.

\section{Spectrally-accurate conformal mapping for smooth boundaries}

\subsection{Computation of the forward map of the boundary to the unit circle}
\begin{figure}
\begin{center}
\scalebox{0.8}{
\def\svgwidth{6.4in}
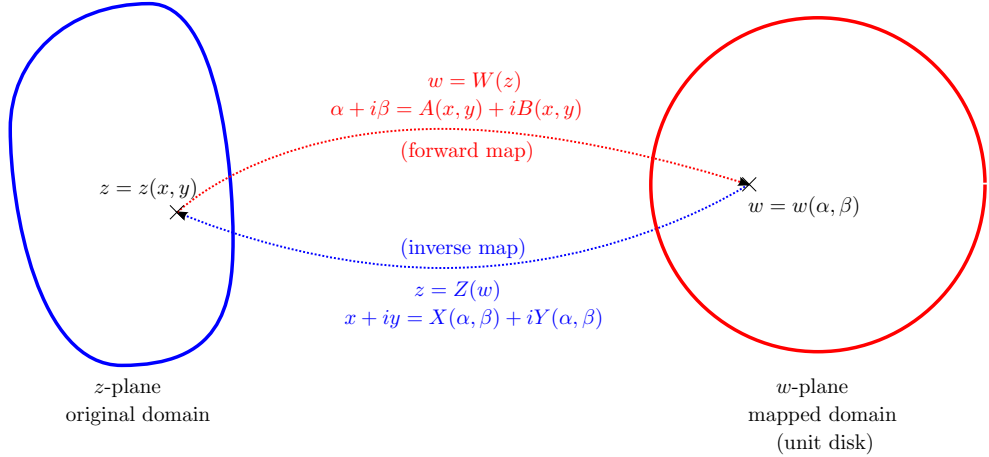
}
\end{center}
\vspace{-8pt}
\caption{Conformal map of a domain to the unit disk.}
\label{fig-ConfMap-Notation}
\end{figure}

In this section, the notation will be as follows.
The original $(R,Z)$ domain will be identified with the complex $z$-plane,
with a point in this plane denoted as $z=x+iy$.
The image domain will be referred to as the $w$-plane,
with a point denoted as $w=\alpha+i\beta$.
The forward map is the complex valued function $w=W(z)$
with real and imaginary parts $\alpha=A(x,y)$ and $\beta=B(x,y)$.
The inverse map is the complex valued function $z=Z(w)$ with real and imaginary parts
$x=X(\alpha,\beta)$ and $y=Y(\alpha,\beta)$
(see Figure \ref{fig-ConfMap-Notation}).
Note that the function $Z$ is obviously not the same mathematical object
as the coordinate $Z$ of the $(R,\phi,Z)$ coordinate system used in this article;
this notation should not lead to any confusion since the two objects are never used simultaneously
and can easily be distinguished based on the context.

The Riemann mapping theorem guarantees the existence and uniqueness 
of an analytic map between any connected set and the unit disk. It is 
straightforward to 
show that under such a mapping, solving the Poisson problem
\[
\lapl u(x,y)=f(x,y), \qquad u\big\vert_{\partial\bOmega} = 0
\]
in the original domain $\bOmega$ is equivalent to solving the 
scaled Poisson problem
\[
\lapl v(\alpha,\beta)
=f(X(\alpha,\beta),Y(\alpha,\beta))\bigg|\frac{dZ}{dw}\bigg|^{2},
\qquad v\big\vert_{\partial\bD_{1}} = 0
\]
on the mapped domain $\bD_1$ (in our case the unit disk),
with
\[
u(x,y) = v(A(x,y), B(x,y)).
\]
We compute the forward map $W$ on $\partial\bOmega$
using the Kerzman-Stein integral equation, 
see \cite{Kerzman_Stein,Kerzman_Trummer}.
In particular, the derivative of the
forward conformal map $W$ on $\partial \Omega$ is given by
\begin{equation}
\begin{split}
	W'(z) &= 2\pi\frac{S^{2}(z,a)}{S(a,a)} ,\\
        S(a,a) &= \int_{\partial\bOmega}\overline{S(z,a)}S(z,a)ds_{z},
	\label{eq:forward_map_deriv}
\end{split}
\end{equation}
where $a$ is the point mapped to the origin, i.e. $W(a) = 0$.
In the examples that follow, 
we always choose $a$ as the geometric center of
$\Omega$~(see Figure \ref{fig:geometry}).
The function $S$ is the Szeg\H o kernel,
which satisfies the integral equation
\begin{equation}\label{eq:Szego_int}
S(z,a)+\int_{\partial\bOmega}\mathcal{A}(z,t) \, S(t,a) \, ds_{t}
=\overline{H(a,z)} .
\end{equation}
Here, the overline represents the complex conjugate, $H$ is the Cauchy kernel,
\[
H(w,z)=\frac{1}{2\pi i} \frac{dz/dl}{z-w} \, ,
\]
and the kernel $\mathcal{A}$ is known as the Kerzman-Stein kernel, defined as
\begin{align*}
	\mathcal{A}(z,w)
        &= \left\{
          \begin{array}{ll}
            \overline{H(w,z)}-H(z,w),&  z\neq w
            \\[6pt]
            0, & z=w \, .
          \end{array}
        \right.
\end{align*}
Numerical implementation of \eqref{eq:Szego_int} is straightforward 
since the kernel $\mathcal{A}$ is smooth.
Since the trapezoidal rule is spectrally-accurate when applied to
sufficiently smooth periodic functions,
the boundary $\partial\bOmega$, and therefore
the integral in \eqref{eq:Szego_int}, 
can be discretized
using $N_{1}$ points equally spaced in arc-length.
The number of discretization points $N_{1}$ is chosen so that
the boundary parameterization and its tangential
derivatives are fully resolved to machine precision.
Letting $z_k$ and $(dz/dl)_k$ denote equispaced points and
tangential derivatives on $\partial\Omega$ (specified either directly,
or obtained from an equation of the plasma surface),
the integral equation in \eqref{eq:Szego_int} is then
discretized resulting in a linear system 
of the form $\mathbf{Ms}=\mathbf{b}$, with
\begin{align*}
	M_{jk} &= \left\{ \begin{array}{ll}
            \displaystyle -\frac{h}{2\pi i}
            \left[\overline{\left[\frac{(dz/dl)_{j}}{z_{j}-z_{k}}\right]} 
              + \frac{(dz/dl)_{k}}{z_{k}-z_{j}}\right],
            & j\neq k \\[12pt]
            1, &j=k \, ,
          \end{array}\right.
        \\[10pt]
        b_{j}
        &=
        -\frac{1}{2\pi i}\overline{\left[\frac{(dz/dl)_{j}}{z_{j}-a}\right]},\\
        s_{k}&=S(z_{k},a),
\end{align*}
where $h=L/N_{1}$ is the step size (with $L$ denoting the length of the 
boundary $\partial\Omega$).
Naive inversion of the linear system $\mathbf{Ms}=\mathbf{b}$ to 
obtain $\mathbf{s}$ requires $O(N_{1}^{3})$ operations.
There exist
fast methods \cite{Odonnell} which can be used to accelerate the solution 
of this linear system.
However, since the conformal map needs to be computed only once 
in our GS solver,
and since the iterative PDE solver is a significantly more expensive step,
we have relied in our experiments on the simpler dense matrix formulation.

Once $\mathbf{s}$ is known, the derivative of 
the forward map for the points on the boundary can be calculated 
using~\eqref{eq:forward_map_deriv}:
\begin{equation}\label{eq:forward_map_deriv_discr}
W'(z_{k})=2\pi\frac{s_{k}^{2}}{s_{a}} \, ,
\qquad s_{a}=h\sum_{k=1}^{N_{1}}\overline{s_{k}}s_{k}.
\end{equation}
Since $\mathbf{s}$ is known at equispaced points in arc length,
the FFT can be used to compute $W(z_{k})$ by Fourier integration.
The integration constant is determined so that the unit disk is 
centered at the origin.
Specifically, three points on the mapped circle can be used to find its 
center, which is then subtracted from the map.
Note that by using Fourier differentiation,
we can also compute $W''$,
which is necessary for the mapping of the second derivatives of functions
defined on the original domain and the mapped domain~\cite{Pataki}. 

\begin{figure}
  \begin{center}
    \includegraphics[width=1.92in,angle=0]{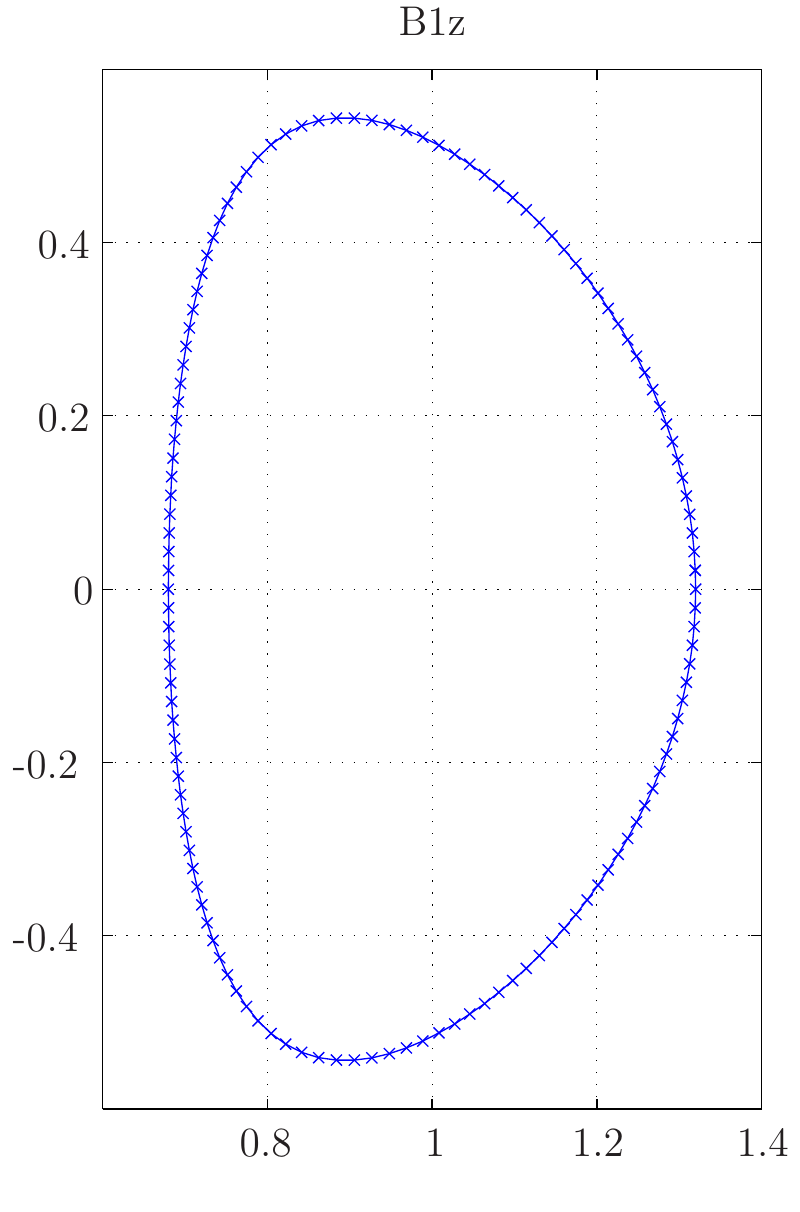}
    \includegraphics[width=2.66in,angle=0]{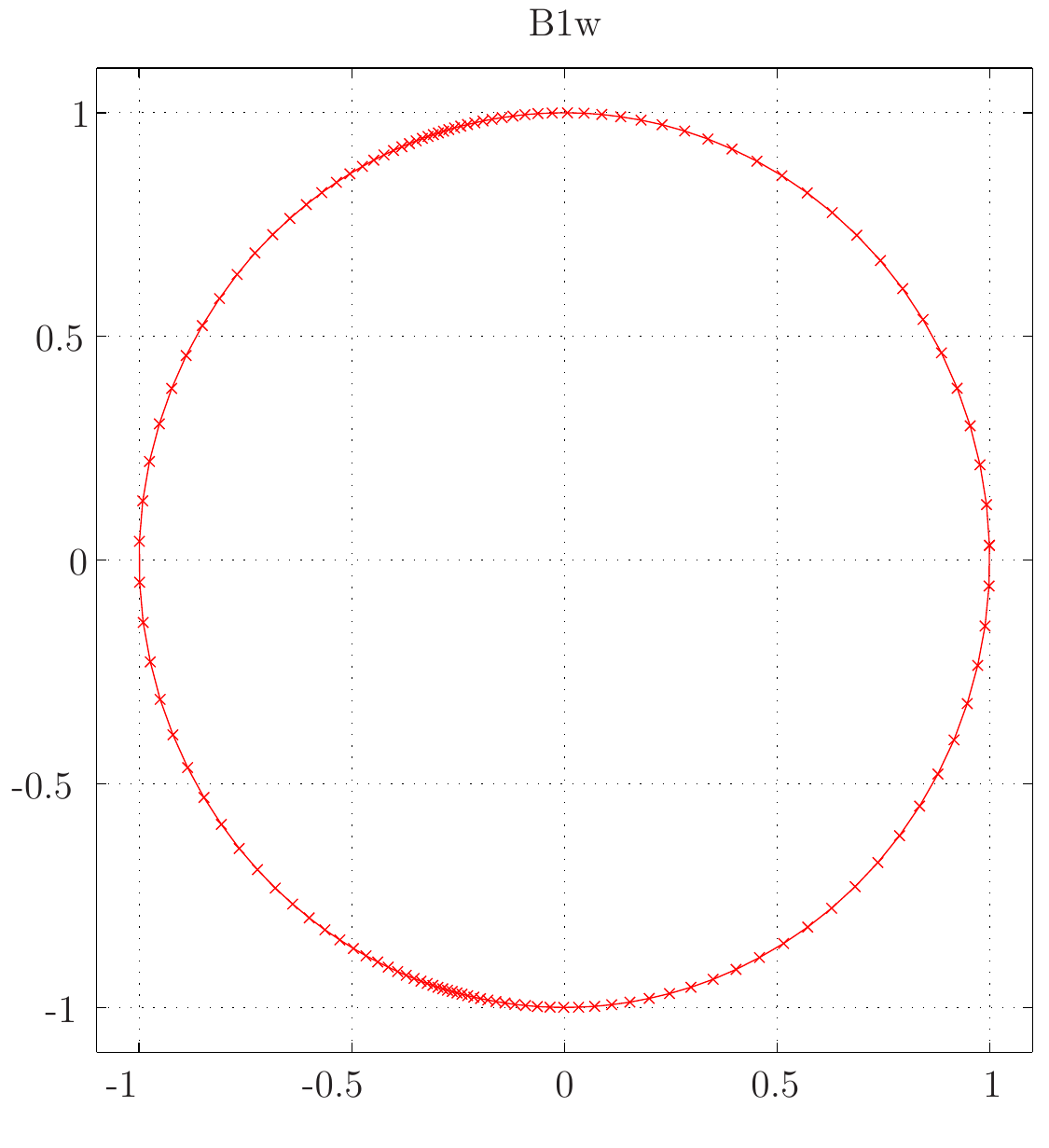}
  \end{center}
  \vspace{-12pt}
  \caption[Conformal map via the Kerzman-Stein integral equation]
    {Conformal map of boundary via the Kerzman-Stein integral equation:
    original domain (left), map of points to the unit circle (right).}
  \label{fig-ConformalMap-Grid1}
\end{figure}

An example of the boundary values of the forward conformal map
is shown in Figure~\ref{fig-ConformalMap-Grid1}.
It is easy to see that for arbitrary boundaries,
equally spaced points on the boundary $\partial\Omega$
do not map to equally spaced points on the unit circle,
a well known feature of conformal mapping known as \emph{crowding}.
In fact, the level of crowding is exponential in 
the aspect ratio of $\Omega$ \cite{Odonnell}.
In many cases, high-order Poisson solvers on the disk which take advantage of
Fourier methods require that
discretization points be equally 
spaced in the angular direction.
Since these points do not coincide with the ones obtained via the
forward conformal map, it is necessary to resample the mapped points.

\subsection{Computing the inverse map for equally spaced points on the unit circle}
\label{sec:conform_interp}
\begin{figure}
  \begin{center}
    \includegraphics[width=1.92in,angle=0]{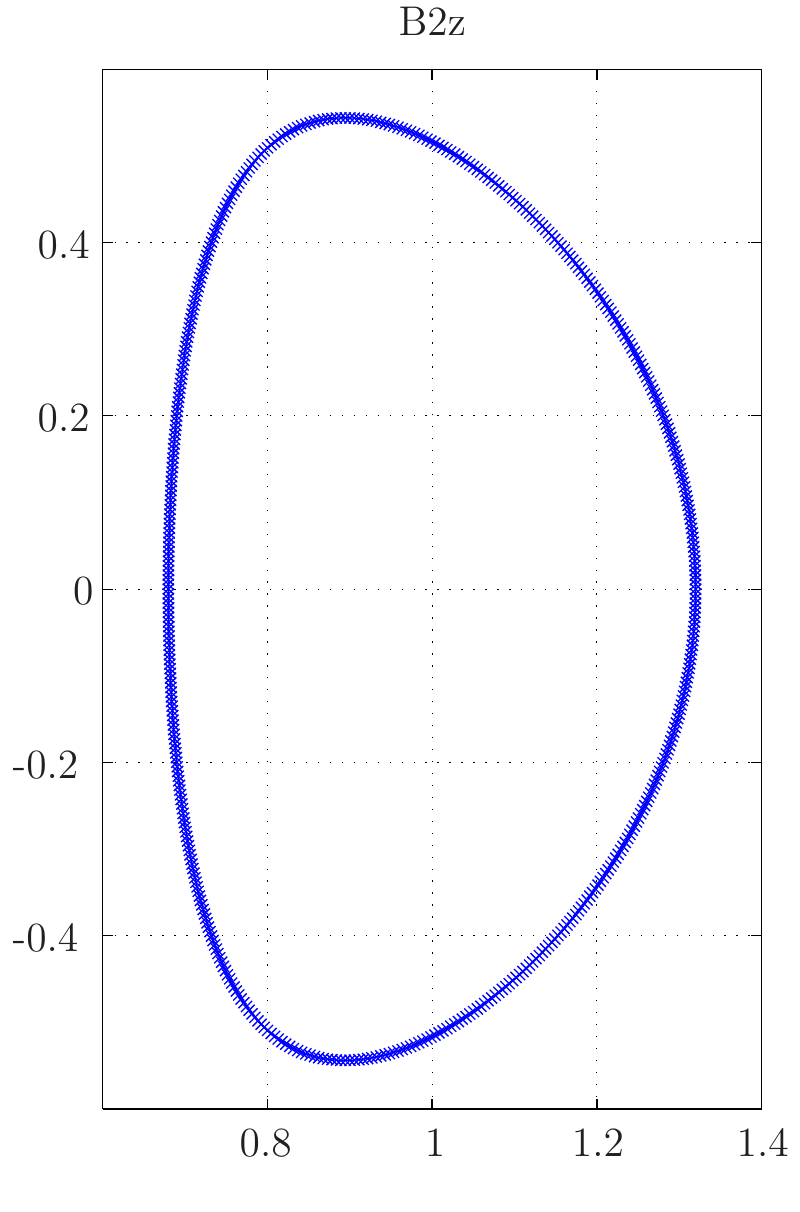}
    \includegraphics[width=2.66in,angle=0]{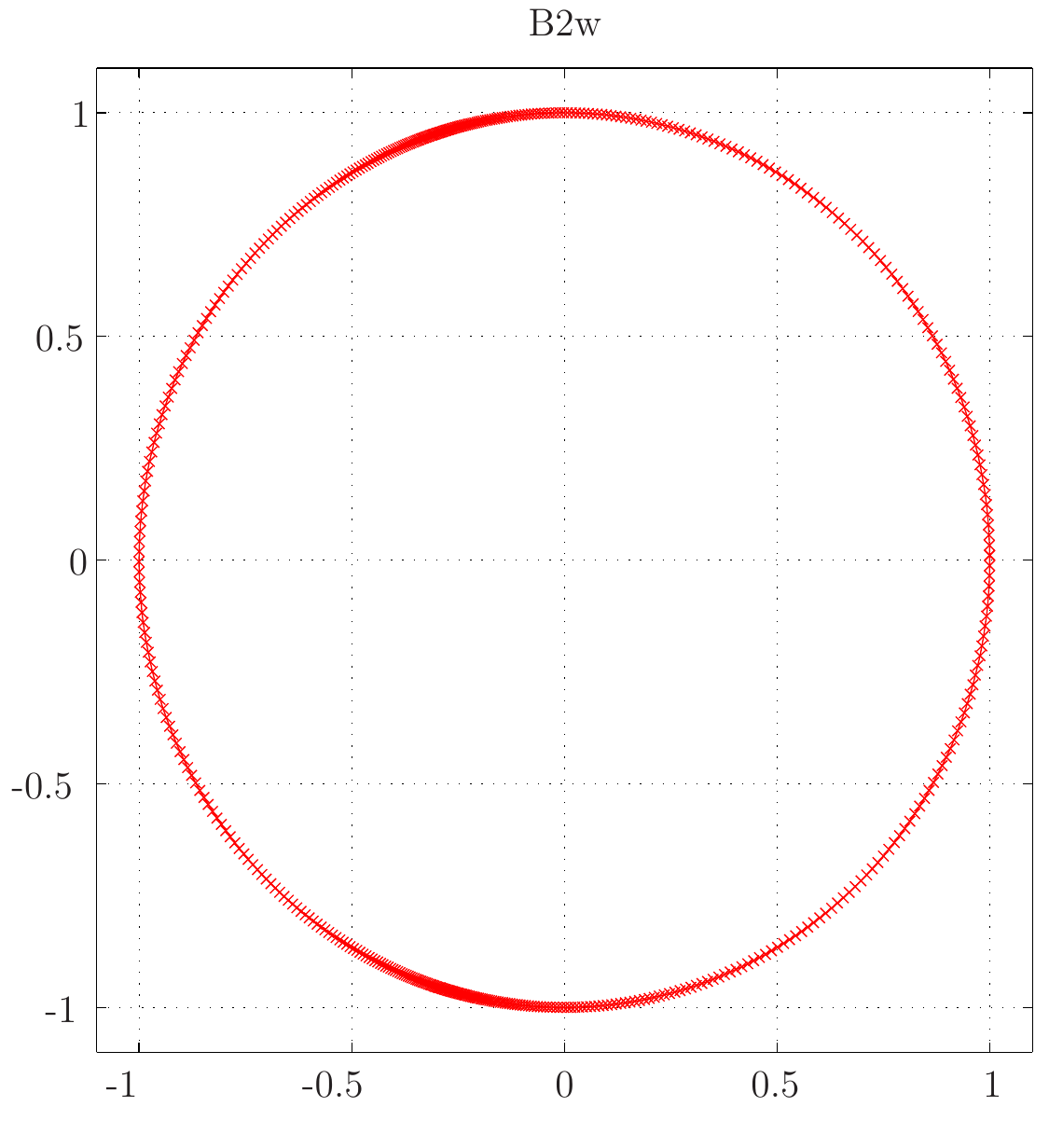}
  \end{center}
  \vspace{-12pt}
  \caption[Conformal map via the Kerzman-Stein integral equation]
    {Conformal map of boundary after oversampling:
    original domain (left), map of points to the unit circle (right).}
  \label{fig-ConformalMap-Grid2}
\end{figure}

To obtain a uniformly spaced angular grid on the unit disk, we 
proceed in two steps.
First, the boundary values of the forward map $W: \Omega \to \bD_1$ 
are oversampled via the FFT.
This is shown in Figure \ref{fig-ConformalMap-Grid2}.
The amount of oversampling necessary is at least the crowding factor of the map. 

\begin{figure}
  \begin{center}
    \includegraphics[width=1.92in,angle=0]{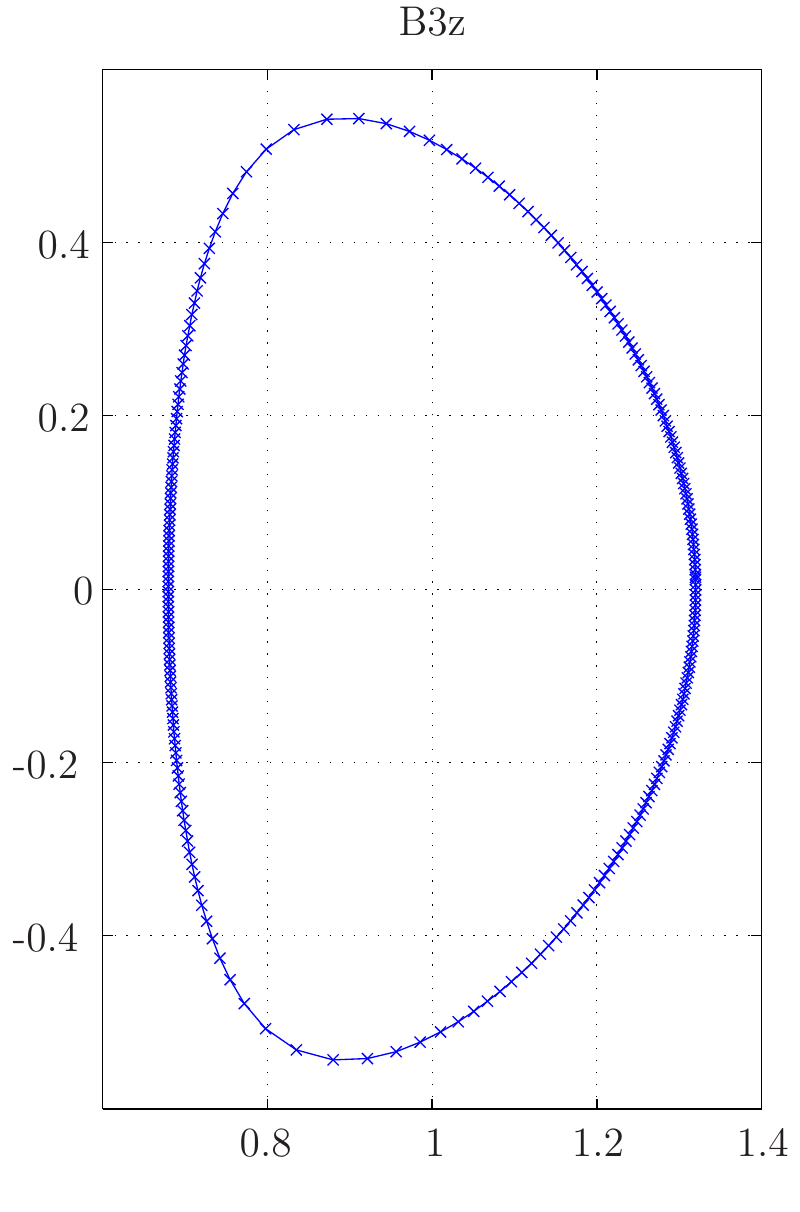}
    \includegraphics[width=2.66in,angle=0]{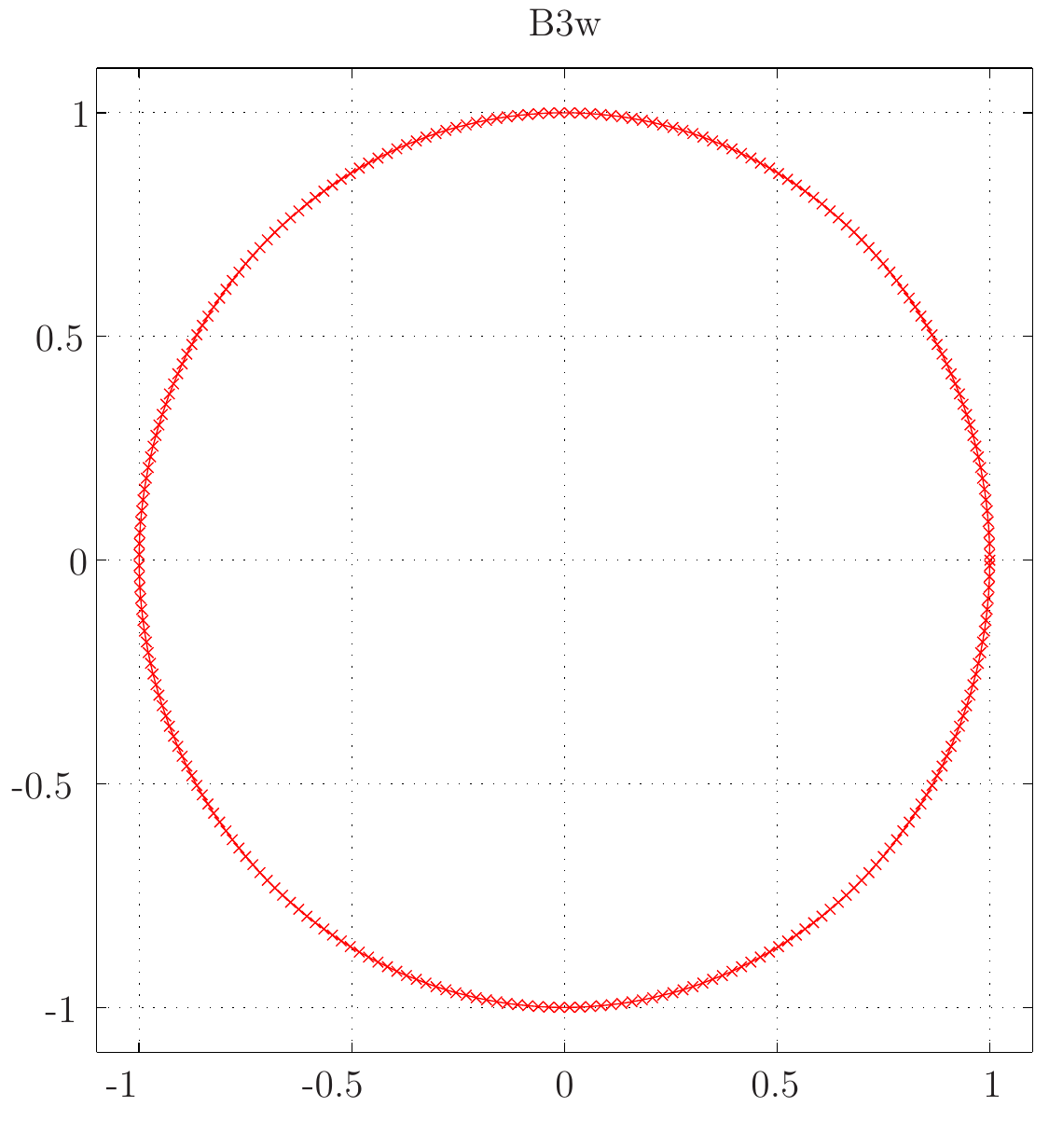}
  \end{center}
  \vspace{-12pt}
  \caption{Inverse of the conformal map of the boundary $-$ equally spaced on the circle: original domain (left), unit circle (right).}
  \label{fig-ConformalMap-Grid3}
\end{figure}

Second, the image points on $\partial\bD_1$ (the unit circle) 
are resampled to equispaced ones,
effectively inverting the conformal map.
We consider $z \in \partial\Omega$ as a function of 
$w \in \partial\bD_1$ and use Lagrange interpolation
to find $z$ values corresponding to equally spaced $w$ values
on the circle.
In particular, barycentric interpolation is used,
which behaves well even for high order~\cite{berrut}.
For example, $8^{\text{th}}$-order barycentric interpolation uses
as data
$4$ points on the left and the right of the target interpolation point.
The oversampling rate in the previous step can be tuned so that 
this interpolation procedure gives the desired accuracy, sometimes requiring
a rate larger than the crowding factor.
The end result of the interpolation step is shown in 
Figure~\ref{fig-ConformalMap-Grid3}. 

The equally spaced $w$ boundary points we obtain in this step will be
used for the angular grid of a high-order Poisson solver. 
We know their image $z=Z(w)$ on the boundary, but do not yet know the 
image of \emph{interior} points of the grid.
To compute the inverse map $Z$ (as well as $Z'$ and $Z''$) for interior points,
the Cauchy integral formula can be used.
Since the boundary of the domain is the unit circle,
the integral formula is particularly easy to evaluate with 
the FFT.
The inverse map of the interior points of a regularly spaced 
$N_{r}\times N_{\theta}$ grid on the unit disk
can be computed in $O(N_{r}N_{\theta}\log N_{\theta})$ time,
the same computational complexity as the Poisson solver on the disk (as
shown in the next section).
This method is spectrally-accurate, and works all the way up to the boundary
without any loss of precision or any requirement for adaptive integration
(due to the proximity of the singularity of the Cauchy integral to the contour,
see~\cite{Pataki} for a detailed discussion).

We end this section with a short discussion on the previously mentioned
issue of \emph{crowding}, inherent to the conformal mapping technique,
and clearly visible in 
Figures~\ref{fig-ConformalMap-Grid1}, \ref{fig-ConformalMap-Grid2},
and~\ref{fig-ConformalMap-Grid3} (see also \cite{Pataki}).
For certain shapes (such as ones with a high aspect ratio), conformal 
maps are clearly not a viable method,
at least when used in combination with solvers which require an equally 
spaced grid on the disk.
For example, a 10 to 1 aspect ratio ellipse has a crowding factor on 
the order of $10^{10}$, which
would obviously lead to prohibitive oversampling.
Fortunately, for the cross-sections of tokamaks and spherical tokamaks,
the crowding factor is much more manageable, on the order of $\sim10$ - $20$.
For numerous geometries relevant to magnetic fusion,
 conformal mapping is an efficient method, as will be shown 
in more detail in Section~\ref{sec:tests}.

\section{High-order Poisson solver on the unit disk}\label{sec:Poisson_disk}

What remains to be discussed in order to complete our Grad-Shafranov solver
is the solution of the two-dimensional Poisson equation
on the unit disk.
There are, of course, a number of fast solvers available for this problem.
For the sake of completeness, we describe one here that is direct, 
high-order accurate, and straightforward to implement. 
Other high-order schemes can be found, for example, in \cite{shen,chen}.

\subsection{Separation of variables and boundary conditions}
Let us consider the fixed-boundary Dirichlet Poisson problem
\begin{align}
  \begin{aligned}
	\lapl v
        &= f
        \hspace{24pt}
        && \mbox{in }\bD_1
        \\
        v
        &= 0
        &&\mbox{on }\partial\bD_1 \, ,
  \end{aligned}
  \label{eq:generic_Poisson}
\end{align}
where $\bD_1$ is the unit disk.
Using separation of variables in the usual polar coordinates
$(r,\theta)$,
we represent both the solution $u$ and the right-hand $f$
side as Fourier series:
\begin{equation}
	v(r,\theta)
        =\sum_{n=-\infty}^{\infty}\hat{v}_{n}(r)e^{in\theta},
        \qquad
        f(r,\theta)
        =\sum_{n=-\infty}^{\infty}\hat{f}_{n}(r)e^{in\theta} .
	\label{eq:Fourier_expansion}
\end{equation}
Substituting these expressions into (\ref{eq:generic_Poisson}),
we have the following radial ordinary differential equation 
for each Fourier mode $n$:
\begin{align}
  \begin{aligned}
	\hat{v}_{n}^{''}(r)+\frac{1}{r}\hat{v}_{n}^{'}(r)-\frac{n^{2}}{r^{2}}\hat{v}_{n}(r)
        &=\hat{f}_{n}(r)
        \\
        \hat{v}_{n}(1)
        &=0 \, .
  \end{aligned}
  \label{eq:radial_Poisson}
\end{align}
In order for equation~(\ref{eq:radial_Poisson}) to be well-posed,
a second boundary condition must be enforced.
We obtain this condition by requiring regularity of 
the solution as $r\rightarrow 0$.
For the $n=0$ mode, multiplying (\ref{eq:radial_Poisson}) by $r$ and
taking the limit $r\rightarrow0$ leads to the condition $\hat{v}_{n}^{'}(0)=0$,
under the assumption that $\hat{f}_{n}$, $\hat{v}_{n}$, and 
$\hat{v}_{n}^{'}$ are bounded.
For $n\neq0$, multiplying the equation by $r^{2}$ and again taking the limit $r\rightarrow0$
leads to $\hat{v}_{n}(0)=0$, under the same boundedness assumptions.
In summary, we are to solve the following decoupled system of equations
in the radial direction:
\begin{equation}
  \begin{aligned}
            \hat{v}_{n}^{''}(r)+\frac{1}{r}\hat{v}_{n}^{'}(r)-\frac{n^{2}}{r^{2}}\hat{v}_{n}(r)=\hat{f}_{n}(r)\;,\;\;\;\hat{v}_{n}^{'}(0)=0\;,\;\;\;\hat{v}_{n}(1)=0\qquad \text{for } n=0\, ,\\
            \hat{v}_{n}^{''}(r)+\frac{1}{r}\hat{v}_{n}^{'}(r)-\frac{n^{2}}{r^{2}}\hat{v}_{n}(r)=\hat{f}_{n}(r)\;,\;\;\;\hat{v}_{n}(0)=0\;,\;\;\;\hat{v}_{n}(1)=0\qquad \text{for } n\neq0 \, .
  \end{aligned}
  \label{eq:radial_Poisson_2BC}
\end{equation}

\subsection{Green's functions solution to the radial equation}
Equation (\ref{eq:radial_Poisson_2BC}) is a well studied ODE in mathematical 
physics, whose Green's function is known.
Using convolution with the Green's function, a 
particular solution $\hat{v}_{n}^{P}$ to the ODE which does 
\emph{not} satisfy the boundary condition at $r=1$ can immediately be
written down.
Then, a correction $\hat{v}_{n}^{H}$ can be found 
which solves the homogeneous equation, and such that 
the sum $\hat{v}_{n}=\hat{v}_{n}^{P}+\hat{v}_{n}^{H}$
satisfies the ODE \emph{and} the boundary conditions,
both at $r=0$ and at $r=1$. To this end, we proceed as follows.

The Green's function 
with the proper behavior at $0$ and infinity
for the ODE in~(\ref{eq:radial_Poisson_2BC}) is
\begin{align*}
	G_{n}(r,s)
        &=\left\{ \begin{aligned}
            s \log s\;,\qquad r<s\\
            s\log r\;,\qquad r>s
          \end{aligned}\right.
        & n&=0 \, ,
        \\
        G_{n}(r,s)
        &=\left\{ \begin{aligned}
            -\frac{1}{2|n|}r^{|n|}s^{-|n|+1}\;,\qquad r<s\\
            -\frac{1}{2|n|}r^{-|n|}s^{|n|+1}\;,\qquad r>s
          \end{aligned}\right. 
        & n&\neq0 \, ,
\end{align*}
and the convolution solutions are
\begin{align*}
	\hat{v}^{P}_{n}(r)
        &= \log r\int_{0}^{r}s\hat{f}_{n}(s)ds+\int_{r}^{\infty}s\log s\hat{f}_{n}(s)ds
        & n&=0 \, ,
        \\
        \hat{v}^{P}_{n}(r)
        &=-\frac{1}{2|n|}\left(r^{-|n|}\int_{0}^{r}s^{|n|+1}\hat{f}_{n}(s)ds+r^{|n|}\int_{r}^{\infty}s^{-|n|+1}\hat{f}_{n}(s)ds\right)
        & n&\neq0 \, .
\end{align*}

Since we are solving the Poisson problem on the unit disk,
we can set $\hat{f}_{n}(r)=0$ for $r>1$, and the convolution 
solutions to the inhomogeneous equation can be written in the form
\begin{subequations}
  \begin{align}
        \label{eq:part_sola}
	\hat{v}^{P}_{n}(r)
        &= \log r\int_{0}^{r}s\hat{f}_{n}(s)ds+\int_{r}^{1}s\log s\hat{f}_{n}(s)ds
        & n&=0 \, ,
        \\
        \label{eq:part_solb}
        \hat{v}^{P}_{n}(r)
        &=-\frac{1}{2|n|}\left(r^{-|n|}\int_{0}^{r}s^{|n|+1}\hat{f}_{n}(s)ds+r^{|n|}\int_{r}^{1}s^{-|n|+1}\hat{f}_{n}(s)ds\right)
        & n&\neq0 \, .
 \end{align}
\end{subequations}

The solution to the homogeneous equation
\begin{equation}\label{eq:radial_Poisson_hom}
\hat{v}_{n}^{''}(r)+\frac{1}{r}\hat{v}_{n}^{'}(r)-\frac{n^{2}}{r^{2}}\hat{v}_{n}(r)=0
\end{equation}
satisfying the regularity condition at $r=0$ is
\begin{align*}
        \hat{v}^{H}_{n}(r)
        &=c_{n}r^{|n|} ,
\end{align*}
where
$c_{n}$ is a constant to be determined from the boundary condition at $r=1$:
\begin{equation*}
	\hat{v}_{n}(1)=\hat{v}^{P}_{n}(1)+\hat{v}^{H}_{n}(1)
        =\hat{v}^{P}_{n}(1)+c_{n}=0 \qquad 
        \Rightarrow \qquad
        c_{n}=-\hat{v}^{P}_{n}(1) \, .
\end{equation*}
Therefore, the general solution to~(\ref{eq:radial_Poisson_2BC}) 
satisfying the boundary conditions at $r=0$ and $r=1$ is
\begin{align}
        \hat{v}_{n}(r)
        &=\hat{v}^{P}_{n}(r)-\hat{v}^{P}_{n}(1)r^{|n|} \, .
  \label{eq:full_sol}
\end{align}
As previously discussed, one of the major advantages of Green's function 
methods is that the radial and angular derivatives 
of the solution $v$ can be calculated explicitly
by differentiating the separation of variables representation.
Numerical differentiation is never required.
This allows for accurate computation of the first and second derivatives
of $\Psi$, which are required for stability and transport calculations.

The radial derivatives are computed by direct differentiation 
of (\ref{eq:part_sola}, \ref{eq:part_solb}) and (\ref{eq:full_sol}).
For the $n=0$ mode, we find
\begin{align}
  \begin{aligned}
	\hat{v}'_{n}(r)
        &=\frac{1}{r}\int_{0}^{r} s \, \hat{f}_{n}(s) \, ds ,
        \\
        \hat{v}''_{n}(r)
        &=\hat{f}_{n}(r)-\frac{1}{r^{2}}\int_{0}^{r}s \, \hat{f}_{n}(s) \, ds .
      \end{aligned}
      \label{eq:n_zero_deriv}
\end{align}
For non-zero $n$ modes, we have
\begin{align}
  \begin{aligned}
    \hat{v}'_{n}(r)
    &=\frac{1}{2} r^{-|n|-1}\int_{0}^{r} s^{|n|+1}\, \hat{f}_{n}(s) \, ds
    - \frac{1}{2} r^{|n|-1}\int_{r}^{1} s^{-|n|+1} \, \hat{f}_{n}(s) \, ds
     - |n|\hat{v}^{P}_{n}(1)r^{|n|-1} ,
    \\
    \hat{v}''_{n}(r) 
    &=\hat{f}_{n}(r)
    - \frac{|n|+1}{2} r^{-|n|-2}\int_{0}^{r}s^{|n|+1}\hat{f}_{n}(s)ds
    \\
    &\hspace{24pt} -\frac{|n|-1}{2} r^{|n|-2}\int_{r}^{1}s^{-|n|+1}\hat{f}_{n}(s)ds
    -|n|(|n|-1)\hat{v}^{P}_{n}(1)r^{|n|-2} .
  \end{aligned}
  \label{eq:n_deriv}
\end{align}

Angular derivatives are obtained via differentiating the Fourier series 
representation (\ref{eq:Fourier_expansion}),
i.e. by a simple multiplication of $in$.
The first and second partial derivatives of the solution $v$ 
are then computed according to
\begin{equation*}
\begin{split}
    v_{r}(r,\theta)=\sum_{n=-\infty}^{\infty}\hat{v}'_{n}(r)e^{in\theta} , &\qquad 
  v_{\theta}(r,\theta)=i\sum_{n=-\infty}^{\infty}n\hat{v}_{n}(r)e^{in\theta},\\
    v_{rr}(r,\theta)=\sum_{n=-\infty}^{\infty}\hat{v}''_{n}(r)e^{in\theta} , &\qquad 
  v_{r\theta}(r,\theta)=i\sum_{n=-\infty}^{\infty}n\hat{v}'_{n}(r)e^{in\theta} , \\
v_{\theta\theta}(r,\theta)=-&\sum_{n=-\infty}^{\infty}n^{2}\hat{v}_{n}(r)e^{in\theta},
\end{split}
\end{equation*}
where the radial derivatives $\hat{v}'_{n}(r)$ and $\hat{v}''_{n}(r)$
are given in equations~(\ref{eq:n_zero_deriv}) and~(\ref{eq:n_deriv}).
As is well known, we observe that taking one
derivative introduces a condition number of $\mathcal O(n)$,
and taking a second derivative introduces a condition number of 
$\mathcal O(n^{2})$, due to the multiplication by $n$ and $n^{2}$ respectively.

\subsection{Numerical considerations}

\subsubsection{Grid setup}

\begin{figure}
  \begin{center}
    \includegraphics[width=1.92in,angle=0]{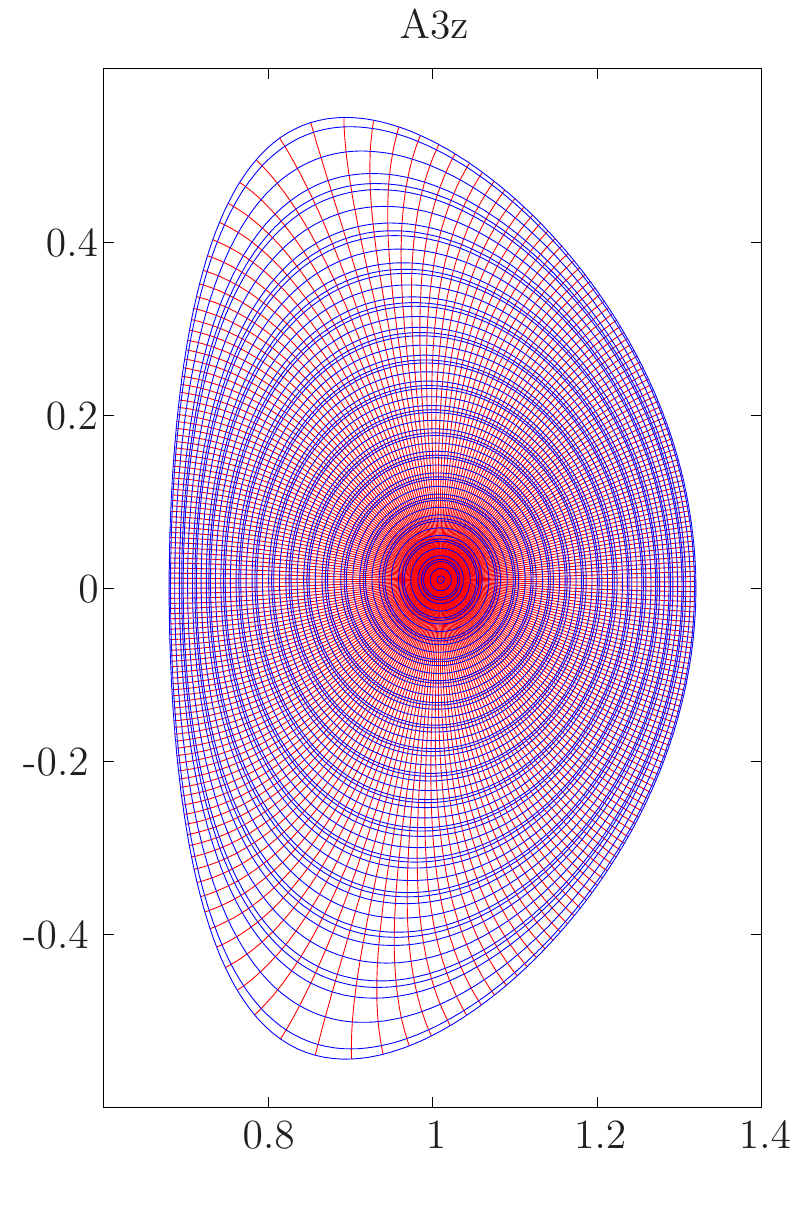}
    \includegraphics[width=2.66in,angle=0]{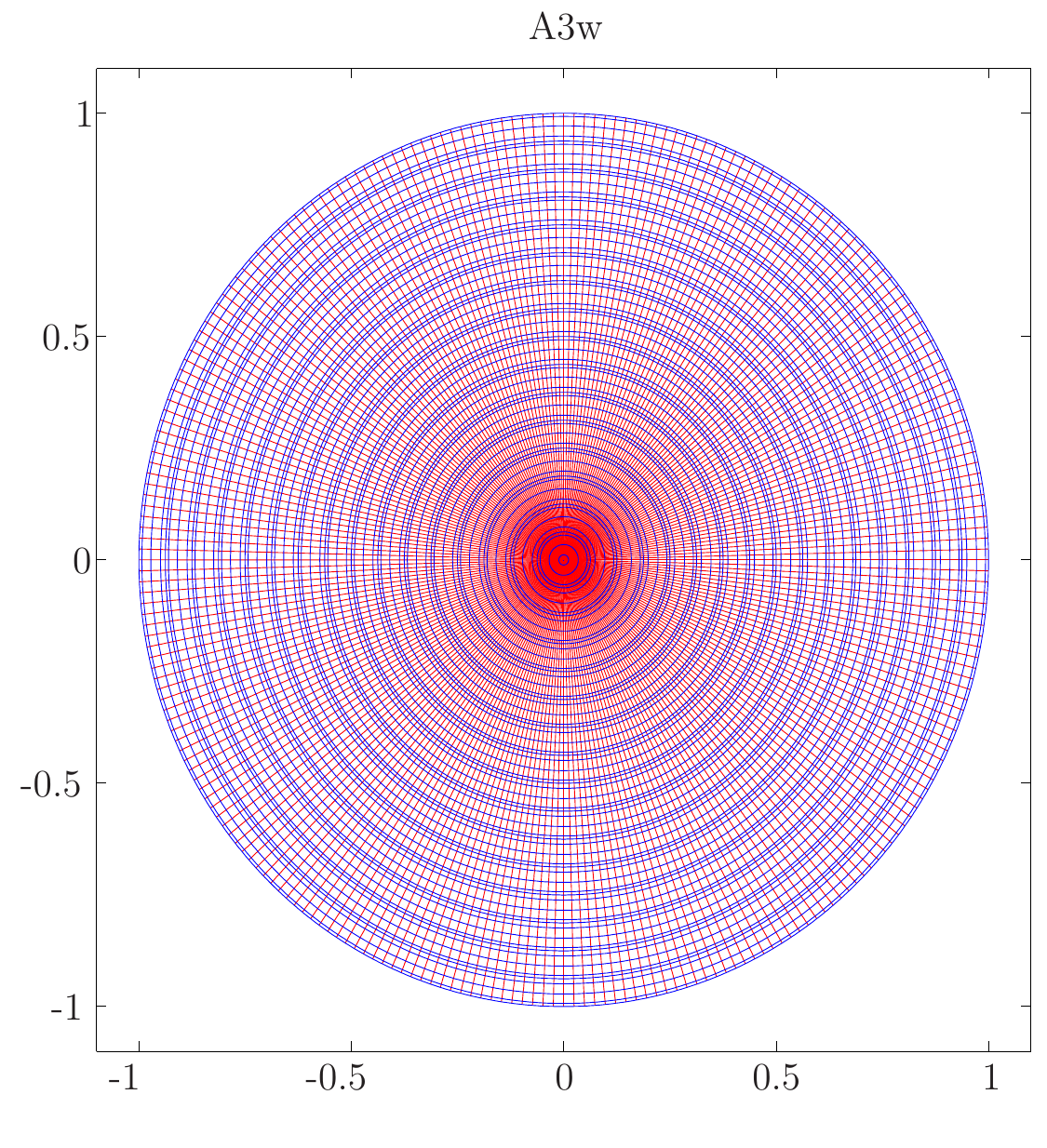}
  \end{center}
  \vspace{-12pt}
  \caption{Tensor grid for the equilibrium calculations in the tokamak presented in Section \ref{sec:tests}. Grid for the unit disk (right) and its image in the original domain (left).}
  \label{fig-ConformalMap-Grid3V}
\end{figure}

Since the Poisson solver previously outlined
relies on separation of variables in polar coordinates,
we build a tensor grid in the radial and angular variables $r$ and $\theta$.
In the $\theta$ variable, we compute and sum Fourier series using the FFT.
When sampled at equispaced points, the representation in terms of Fourier
series is spectrally-accurate for smooth data.
Let $N_{\theta}$ denote the number of equispaced angular grid points 
in $\theta$.
In the radial direction, however, the domain is not periodic.
Therefore, an appropriate high-order representation is 
a piecewise Chebyshev grid:
the interval [0,1] is divided into $N_{L}$ intervals, 
and a $P^{\text{th}}$ order Chebyshev grid is constructed on each interval.
The total number of points in the radial direction is therefore $N_{r}=PN_{L}$.
The order $P$ can be chosen as desired; in all subsequent examples
we set $P=16$, yielding a $16^{\text{th}}$ order scheme.
The grid used for a typical tokamak geometry (discussed in more 
detail in Section~\ref{sec:tests})
is shown in Figure~\ref{fig-ConformalMap-Grid3V},
along with its image under the inverse conformal 
map in the original domain.

\subsubsection{Evaluation of Green's function convolutions}

Naive implementations of formulas~(\ref{eq:part_sola}, \ref{eq:part_solb})
require $O(N_{r}^{2})$ work.
However, the first integrals in~(\ref{eq:part_sola}, \ref{eq:part_solb})
can be computed recursively from $r=0$ to $r=1$, while the 
second integrals in~(\ref{eq:part_sola}, \ref{eq:part_solb}) can be 
computed recursively from $r=1$ to $r=0$. 
Thus, all integrals can be calculated in $O(N_r)$ work.

Furthermore, some care must be taken in the computation of the integrals 
in~(\ref{eq:part_sola}, \ref{eq:part_solb}) near $r=0$
because of the rapid growth/decay of the monomials $s^{|n|}$ and $s^{-|n|}$
for large mode numbers $n$. The difficulties associated with floating-point
overlow/underflow are easily avoided by rescaling. A second problem 
is that for $n$ large, the integrands $r^{|n|}/s^{(|n|-1)} \, \hat{f}_n(s)$ and 
$s^{|n|+1}/r^{|n|} \, \hat{f}_n(s)$ are poorly resolved by the 
composite Chebyshev grid. Since $\hat{f}_n(s)$ is well-resolved, however,
the change of variables $\sigma = (r/s)^{|n|}$ and $\mu = (r/s)^{1-|n|}$ 
for the first and second cases, respectively, yield well-resolved functions
in the transformed variable. These mapped integrals can be computed 
accurately with standard quadrature rules \cite{Pataki}.  In our 
experiments below, we used a $16^{\rm th}$ order Gaussian rule.

\subsubsection{Convergence and run time}

If the data given is smooth, the Fourier series representation is
spectrally-accurate.
The overall order of convergence of the algorithm is therefore $P$, 
the order of the piecewise Chebyshev polynomials used in the radial ODE solver.

The run time complexity of the algorithm is
$\mathcal O(N_{r}N_{\theta}\log N_{\theta})$, 
nearly optimal with respect to the number of grid points.
In detail, we compute $\mathcal O(N_{r})$ FFTs of size $N_{\theta}$ at 
a cost of $\mathcal O(N_{r}N_{\theta}\log N_{\theta})$, and $N_{\theta}$ ODE 
solves of complexity $\mathcal O(N_{r})$ are performed
at a total cost $\mathcal O(N_{r}N_{\theta})$. Note 
that here we treat $P$ as a fixed constant that would not be increased 
if the grid is refined.

\section{Numerical tests -- Examples}\label{sec:tests}
\subsection{Comparison with exact analytic solutions}
\begin{figure}
\begin{minipage}{0.60\linewidth}
\centering
\includegraphics[width=\textwidth]{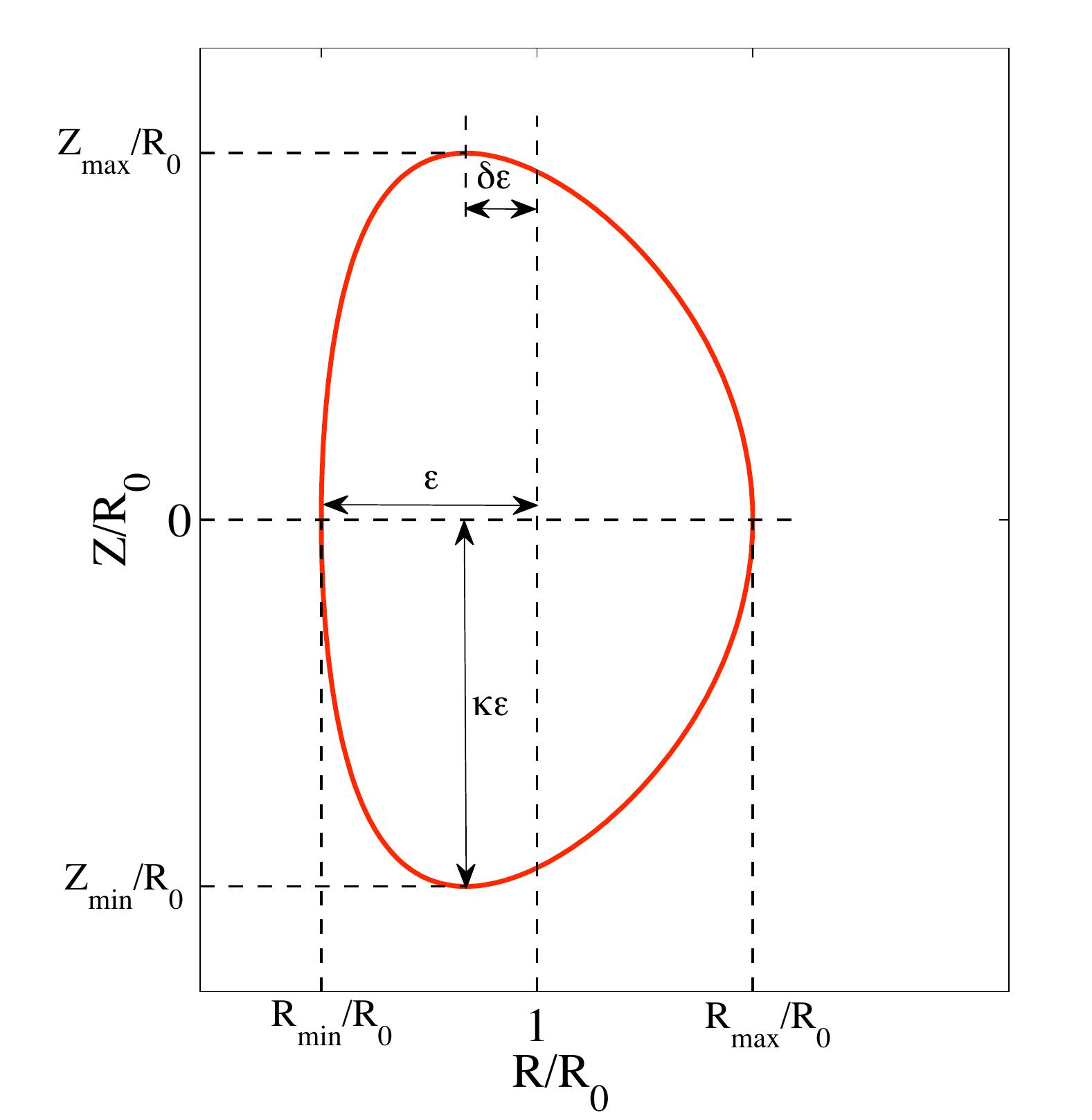}
\end{minipage}
\scalebox{0.9}{
\begin{minipage}{0.40\linewidth}
\begin{align*}
&R_{0}=\frac{R_{max}+R_{min}}{2}\\
&\epsilon=\frac{R_{max}-R_{0}}{R_{0}}=\frac{R_{min}-R_{0}}{R_{0}}\\
&\kappa\epsilon=\frac{Z_{max}}{R_{0}}\\
&1-\delta\epsilon=\frac{R}{R_{0}}\Biggr\vert_{Z=Z_{max}}
\end{align*}
\end{minipage}
}
\caption{Geometric definition of the parameters $\epsilon$, $\kappa$, 
and $\delta$.}
\label{fig:geometry}
\end{figure}

In order to test the accuracy of our Grad-Shafranov solver,
we first consider a case where exact solutions are known.
As discussed in Section \ref{sec:GS_presentation}, these are easy to
construct for profiles of the form given in 
equation~(\ref{eq:profiles_with_constant}) with $S=0$ and $T=0$.
For simplicity we set $A=0$ (corresponding to a plasma which 
is neither paramagnetic nor diamagnetic~\cite{Cerfon})
and normalize the pressure such that $C=1$.
These choices are equivalent to solving the GS equation
\begin{equation}\label{eq:GS_exact_simple}
	\lapl^*\Psi=R^{2},
\end{equation}
which we solve both numerically and analytically.
A simple analytic solution to~(\ref{eq:GS_exact_simple})
which is relevant to magnetic fusion
can be constructed by following the methodology given in~\cite{Cerfon}.
We sum a particular solution to the equation, $R^{4}/8$, 
with three solutions to the homogeneous equation:
\begin{equation}\label{eq:psi_part_1}
\Psi_{exact}(R,Z)=\frac{R^{4}}{8}+d_{1}+d_{2}R^{2}+d_{3}(R^{4}-4R^{2}Z^{2}).
\end{equation}
The free coefficients $d_{1}$, $d_{2}$, and $d_{3}$ are determined so that the contour $\Psi=0$ represents a reasonable plasma cross-section. Specifically, we introduce three characteristic quantities describing the cross-section of a magnetic confinement device, as shown in Figure \ref{fig:geometry}: the inverse aspect ratio $\epsilon$, the elongation $\kappa$, and the triangularity $\delta$. 

\begin{figure}
\begin{center}
\includegraphics[width=4.5in,angle=0]{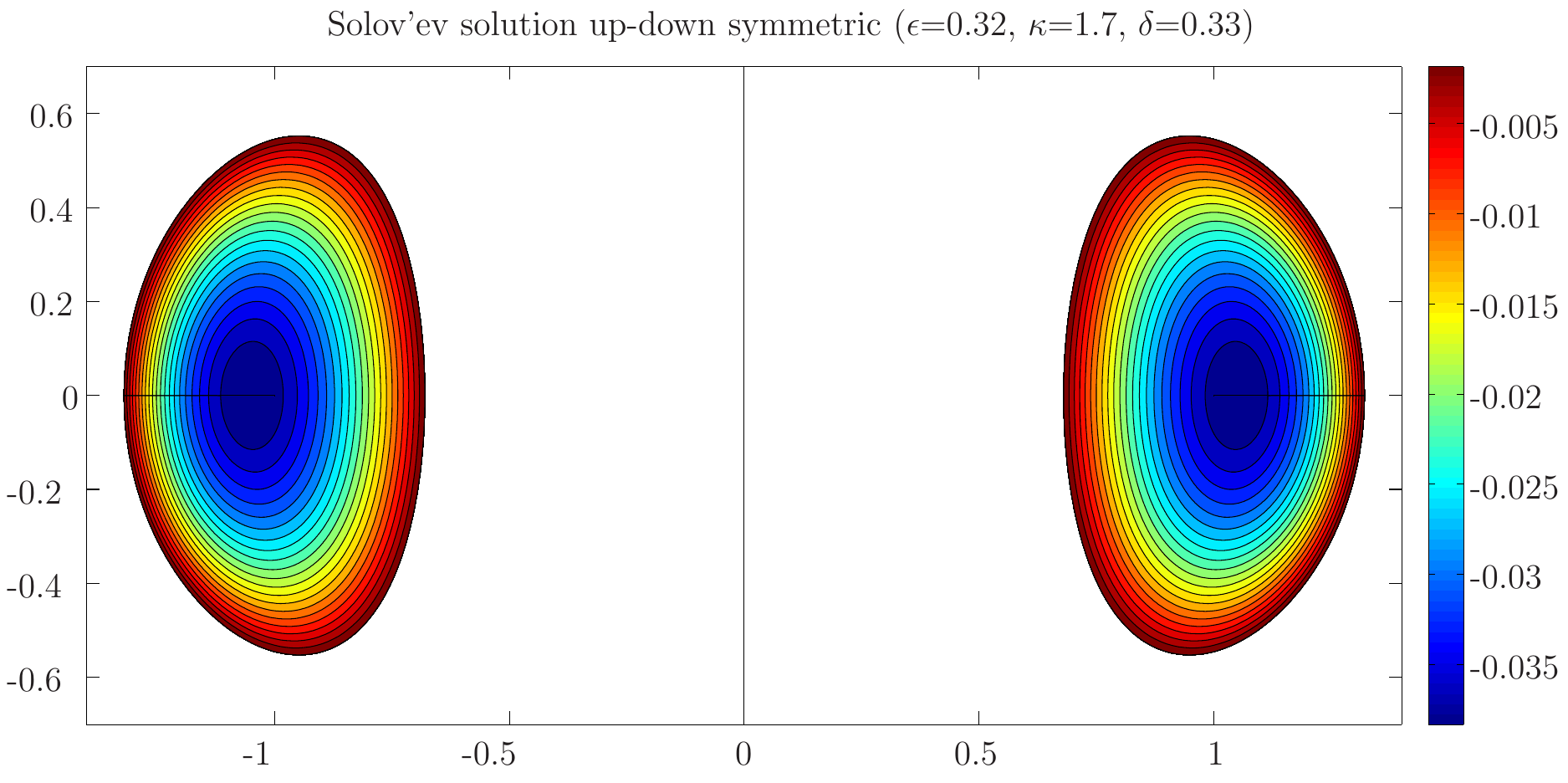}
\end{center}
\vspace{0pt}
\caption{Solution to (\ref{eq:GS_exact_simple}) -- ITER parameters.}
\label{fig-GS-Soloviev-1}
\end{figure}

\begin{figure}
\begin{center}
\includegraphics[width=3.8in,angle=0]{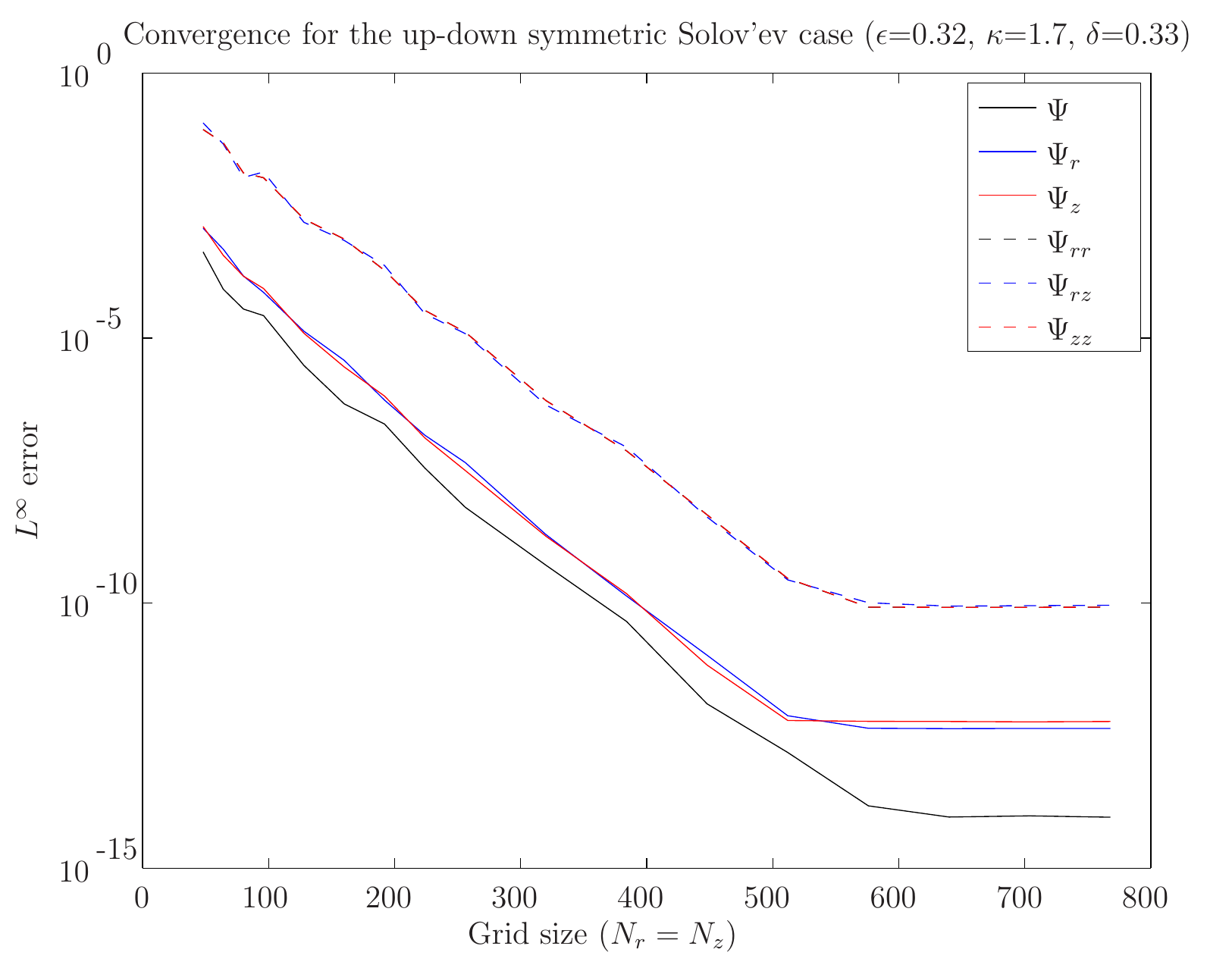}
\end{center}
\vspace{-6pt}
\caption{Convergence of the numerical solution to (\ref{eq:GS_exact_simple}),
and of its first and second derivatives -- ITER parameters.}
\label{fig-GS-Soloviev-3}
\end{figure}

\begin{figure}
\begin{center}
\includegraphics[width=4.0in,angle=0]{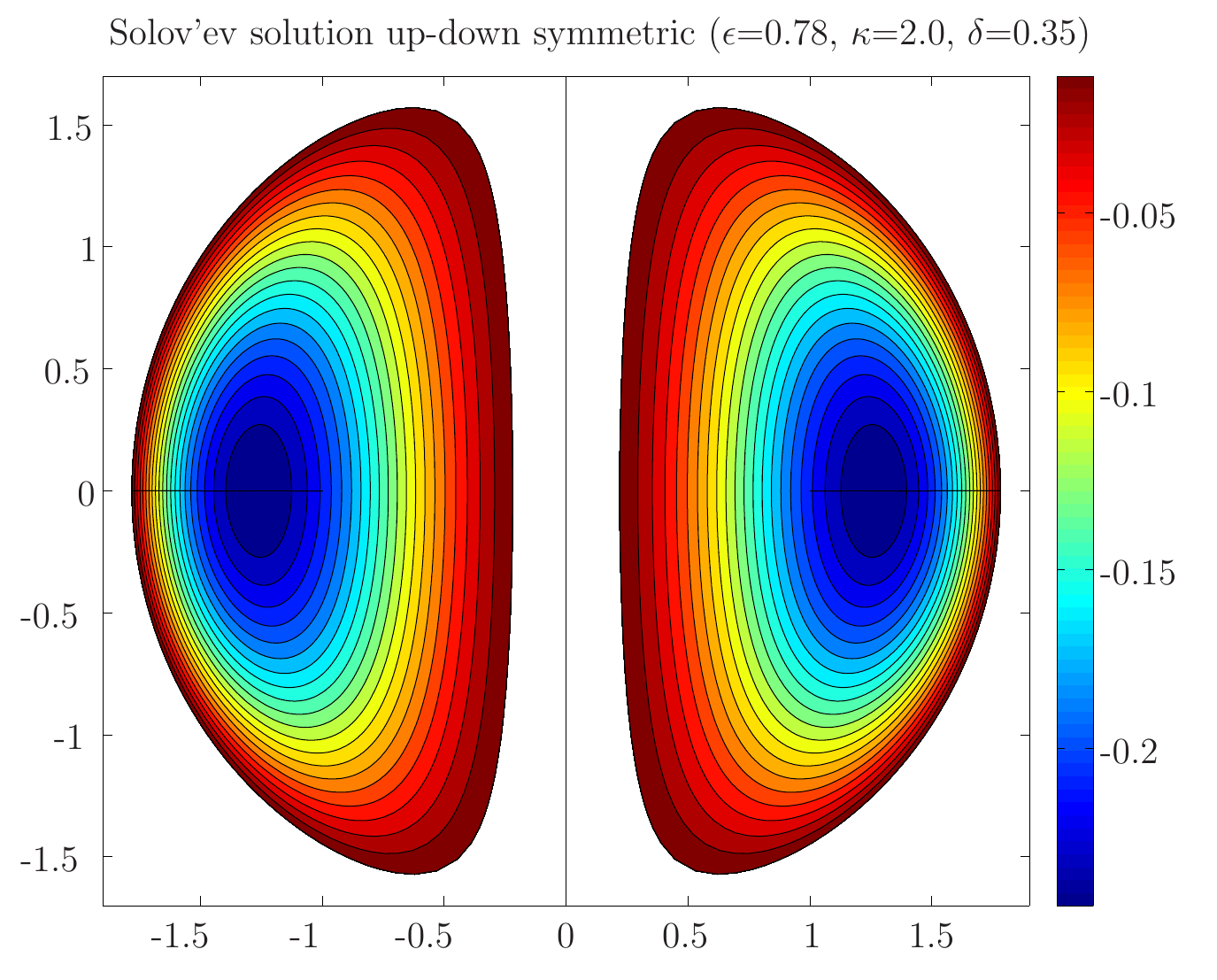}
\end{center}
\vspace{-6pt}
\caption{Numerical solution to (\ref{eq:GS_exact_simple}) -- NSTX parameters.}
\label{fig-GS-Soloviev-2}
\end{figure}

\begin{figure}
\begin{center}
\includegraphics[width=3.8in,angle=0]{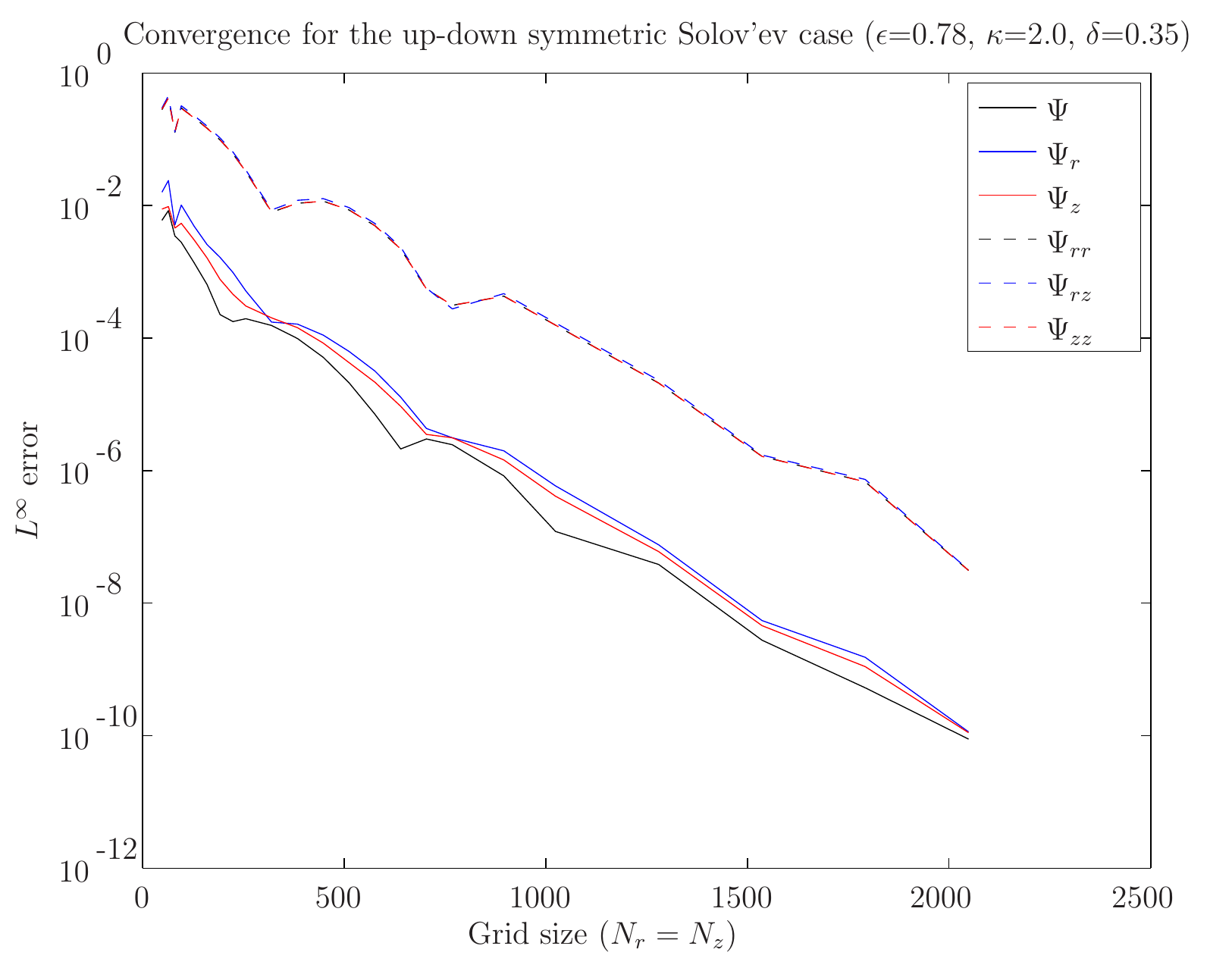}
\end{center}
\vspace{-6pt}
\caption{Convergence of the numerical solution to~(\ref{eq:GS_exact_simple}),
and of its first and second derivatives -- NSTX parameters.}
\label{fig-GS-Soloviev-4}
\end{figure}

The boundary conditions to be enforced are:
\begin{align}
  \begin{aligned}
    \Psi(1+\epsilon,0)
    &=0
    && \mbox{Condition at the outboard midplane,}
    \\
    \Psi(1-\epsilon,0)
    &=0
    && \mbox{Condition at the inboard midplane,}
    \\
    \Psi(1-\delta\epsilon,\kappa\epsilon)
    &=0
    && \mbox{Condition at the top/bottom,}
  \end{aligned}
  \label{eq:BC}
\end{align}
which gives the following system of three equations for $d_{1}$, $d_{2}$ and $d_{3}$:
\begin{align}
\label{eq:system}
	\left[
          \begin{array}{ccc}
            1&(1+\epsilon)^{2}&(1+\epsilon)^{4}\\[4pt]
            1&(1-\epsilon)^{2}&(1-\epsilon)^{4}\\[4pt]
            1&(1-\delta\epsilon)^{2}&(1-\delta\epsilon)^{4}-4(1-\delta\epsilon)^{2}\kappa^{2}\epsilon^{2}
          \end{array}
        \right]
        \left[\begin{array}{c}
            d_{1}\\[4pt]
            d_{2}\\[4pt]
            d_{3}
          \end{array}
        \right]
        &=
        -\frac{1}{8}\left[\begin{array}{c}
            (1+\epsilon)^{4}\\[4pt]
            (1-\epsilon)^{4}\\[4pt]
            (1-\delta\epsilon)^{4}
          \end{array}
        \right].
\end{align}
Equation (\ref{eq:system}) is easily inverted, 
and once the coefficients $d_1, d_2$, and $d_3$ are determined,
the analytic solution to (\ref{eq:GS_exact_simple})
given by (\ref{eq:psi_part_1}) is straightforwardly computed.
Furthermore, the boundary of the plasma is given by the equation
\begin{align*}
	\frac{R^{4}}{8}+d_{1}+d_{2}R^{2}+d_{3}(R^{4}-4R^{2}Z^{2})
        &= 0.
\end{align*}
Using this plasma boundary allows the numerically obtained  solution to 
be compared with the exact analytic one. The grid can then be refined
enabling us to check the convergence of the scheme.
Two examples are shown here:  an ITER-like case \cite{ITER1,ITER2}
with parameters $\epsilon=0.32$, $\kappa=1.7$, $\delta=0.33$,
and an NSTX-like case \cite{NSTX}
with $\epsilon=0.78$, $\kappa=2$, $\delta=0.35$.
Convergence is measured by computing the $L^{\infty}$ norm  (i.e. sup norm)
of the difference between the numerical solution and the exact solution.
For the ITER-like case, contour plots of the numerically obtained 
solution are shown in Figure~\ref{fig-GS-Soloviev-1},
and the convergence behavioris shown in Figure~\ref{fig-GS-Soloviev-3}.
For the NSTX-like case, the analogous figures are~\ref{fig-GS-Soloviev-2} 
and~\ref{fig-GS-Soloviev-4}.

Figures~\ref{fig-GS-Soloviev-3} and~\ref{fig-GS-Soloviev-4} demonstrate
the spectral convergence of the solution and its first and second derivatives.
In the ITER-like case, $\Psi$ is computed with an accuracy close 
to machine precision for a $600$ by $600$ grid.
One can notice that the errors are a bit larger for the first and 
second derivatives.
This is because we solve the Grad-Shafranov equation with 
Dirichlet boundary conditions,
and in order to satisfy the differential equation near the boundary
we effectively compute two derivatives of the boundary condition. 
This procedure introduces numerical differentiation errors, albeit only
of size $\mathcal O(n)$ and $\mathcal O(n^2)$.

One can also observe that the NSTX-like case has worse convergence 
than the ITER-like case, and requires a finer grid to obtain 
comparable precision.
This is a direct consequence of the conformal mapping method.
Indeed, the NSTX-like cross-section has a crowding factor of about 20,
compared to about 7 for the ITER-like case.
Therefore, an extra oversampling factor of 3 is required in constructing
the conformal map.
In addition, the NSTX-like boundary requires slightly more points to 
be resolved.

The GS solver described is obviously not limited to problems 
with up-down symmetry (symmetry about the line $Z=0$).
Similar numerical tests were done with exact up-down asymmetric 
solutions to~(\ref{eq:GS_exact_simple}),
and showed very similar performance~\cite{Pataki}.
Not suprisingly, the solver is also not limited to low pressure 
cases (low-$\beta$).
Equilibria with significant Shafranov shifts have been computed 
without any difficulty~\cite{Pataki}.

\subsection{Numerical equilibrium with pressure pedestal}

As an illustration of the Grad-Shafranov equation as a nonlinear eigenvalue problem (see Section \ref{sec:GS_presentation}), we consider a generic pressure profile corresponding to an equilibrium with a pressure pedestal \cite{Beurskens_ped,Maddison_ped}:
\begin{equation*}
p(\Psi)=(C_{1}+C_{2}\Psi^{2})\left(1-e^{-\Psi^{2}/\eta}\right),
\end{equation*}
where $C_{1}$ and $C_{2}$ are normalization constants,
set to $C_{1}=0.8$ and $C_{2}=0.2$ in the examples shown below.
The parameter $\eta$ is a constant associated with the width and 
the steepness of the pressure pedestal.
Equilibria are calculated for three different pedestal steepnesses 
corresponding to
$\eta=0.1$, $\eta=0.02$ and $\eta=0.005$ (see Figure~\ref{fig-GS-Pedestal-1}).
Assuming, as before, that the plasma is neither paramagnetic nor 
diamagnetic ($dg^{2}/d\Psi = 0$),
the GS equation becomes the following nonlinear eigenvalue problem
(to within a constant factor that is subsumed into the eigenvalue):

\begin{figure}
\begin{center}
\includegraphics[width=2.5in,angle=0]{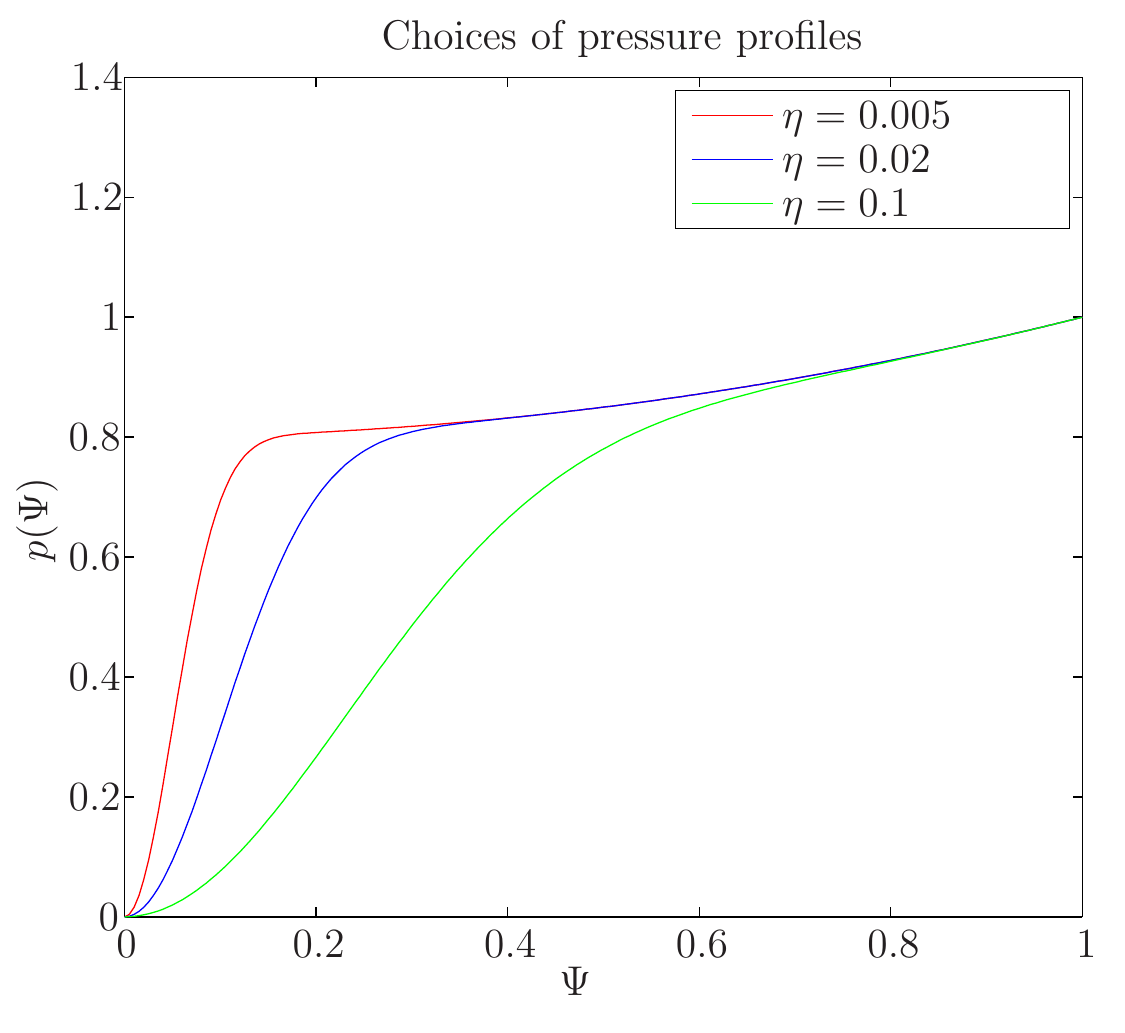}
\includegraphics[width=2.5in,angle=0]{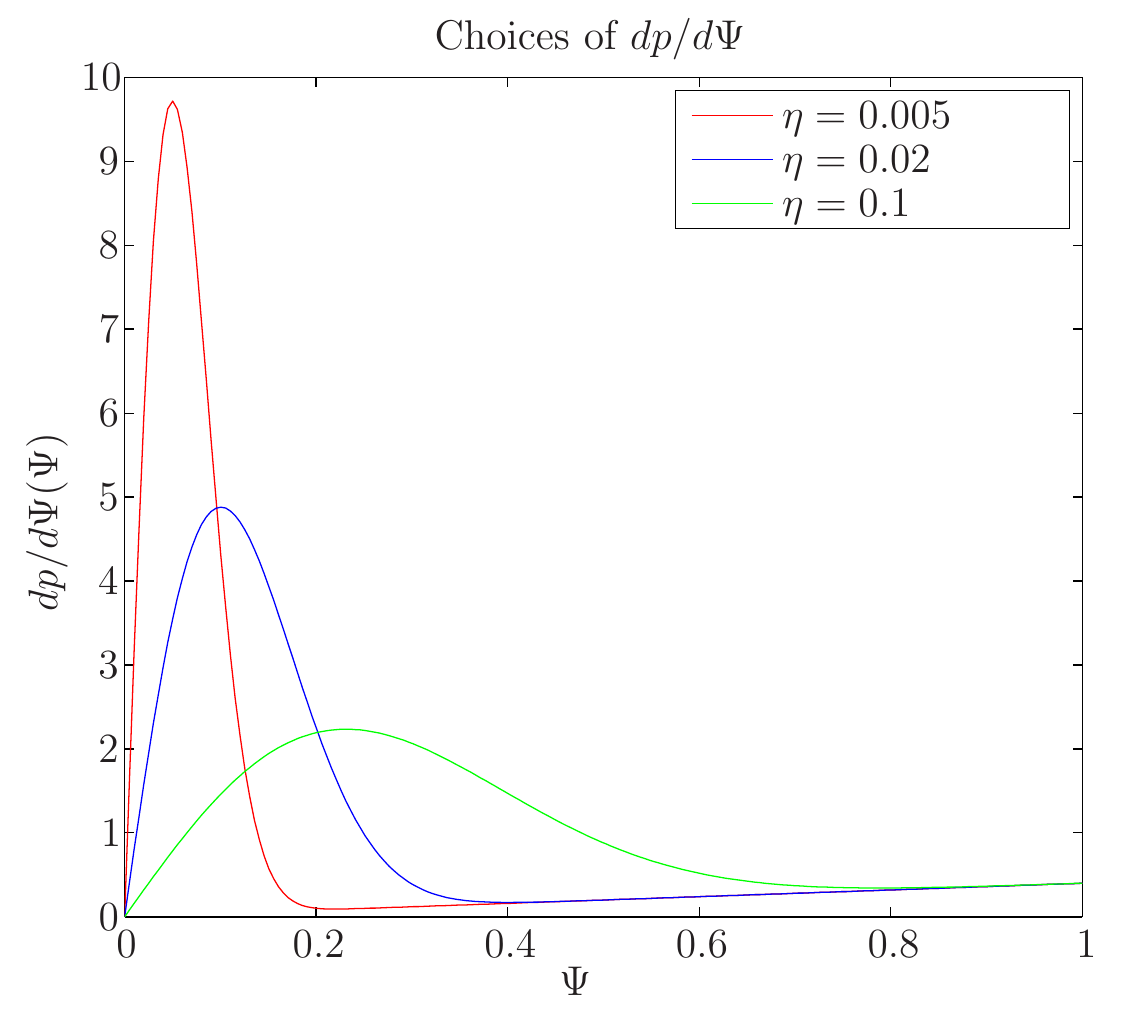}
\end{center}
\vspace{-6pt}
\caption{Pressure profiles for $\eta = 0.005,\, 0.02,\, 0.1$.}
\label{fig-GS-Pedestal-1}
\end{figure}

\begin{figure}
\begin{center}
\includegraphics[width=5.0in,angle=0]{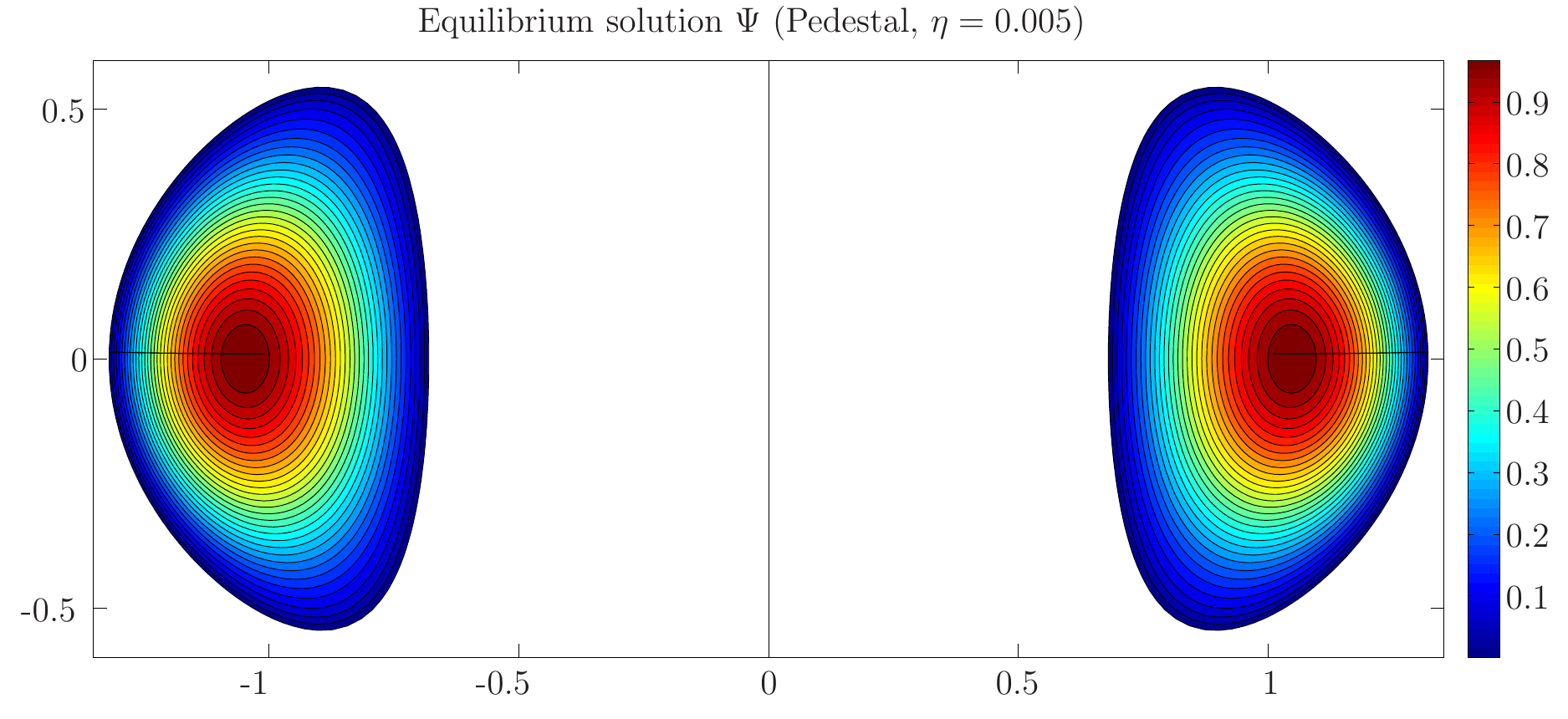}
\end{center}
\vspace{0pt}
\caption{Numerical solution to (\ref{eq:GS_eigen_ex}) for $\eta=0.005$.}
\label{fig-GS-Pedestal-2}
\end{figure}

\begin{figure}
\begin{center}
\includegraphics[width=4.5in,angle=0]{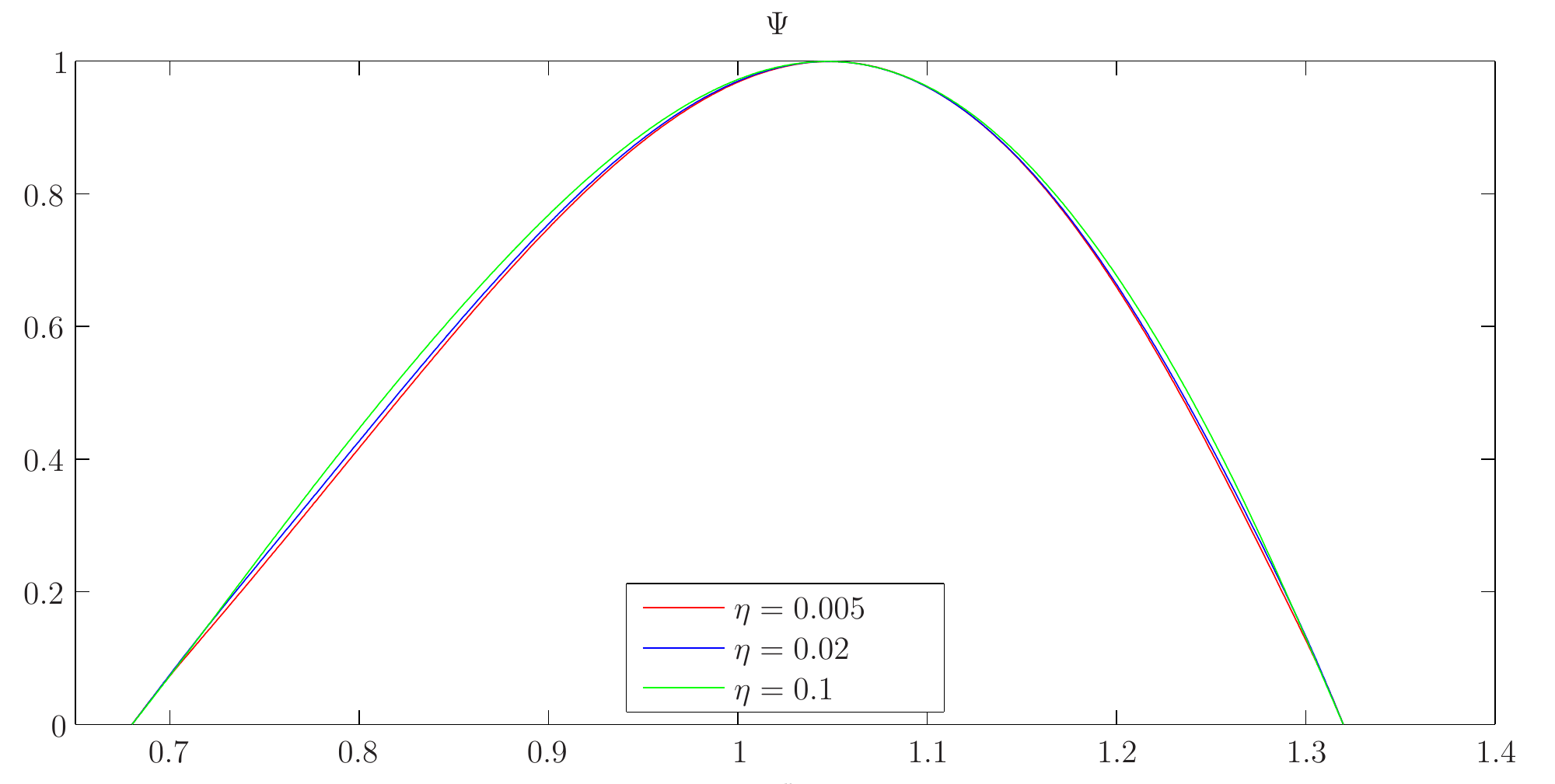}
\includegraphics[width=4.5in,angle=0]{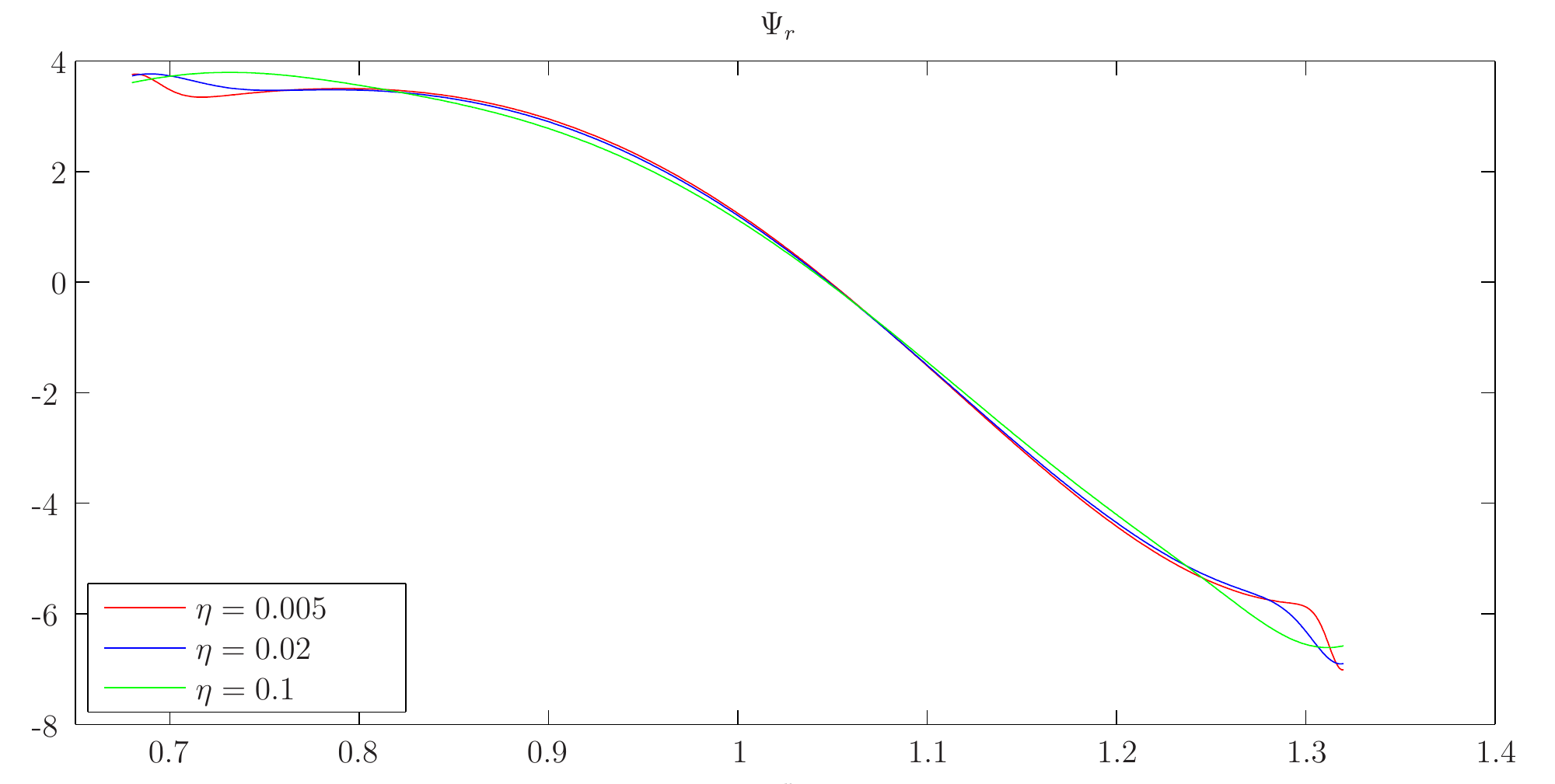}
\includegraphics[width=4.5in,angle=0]{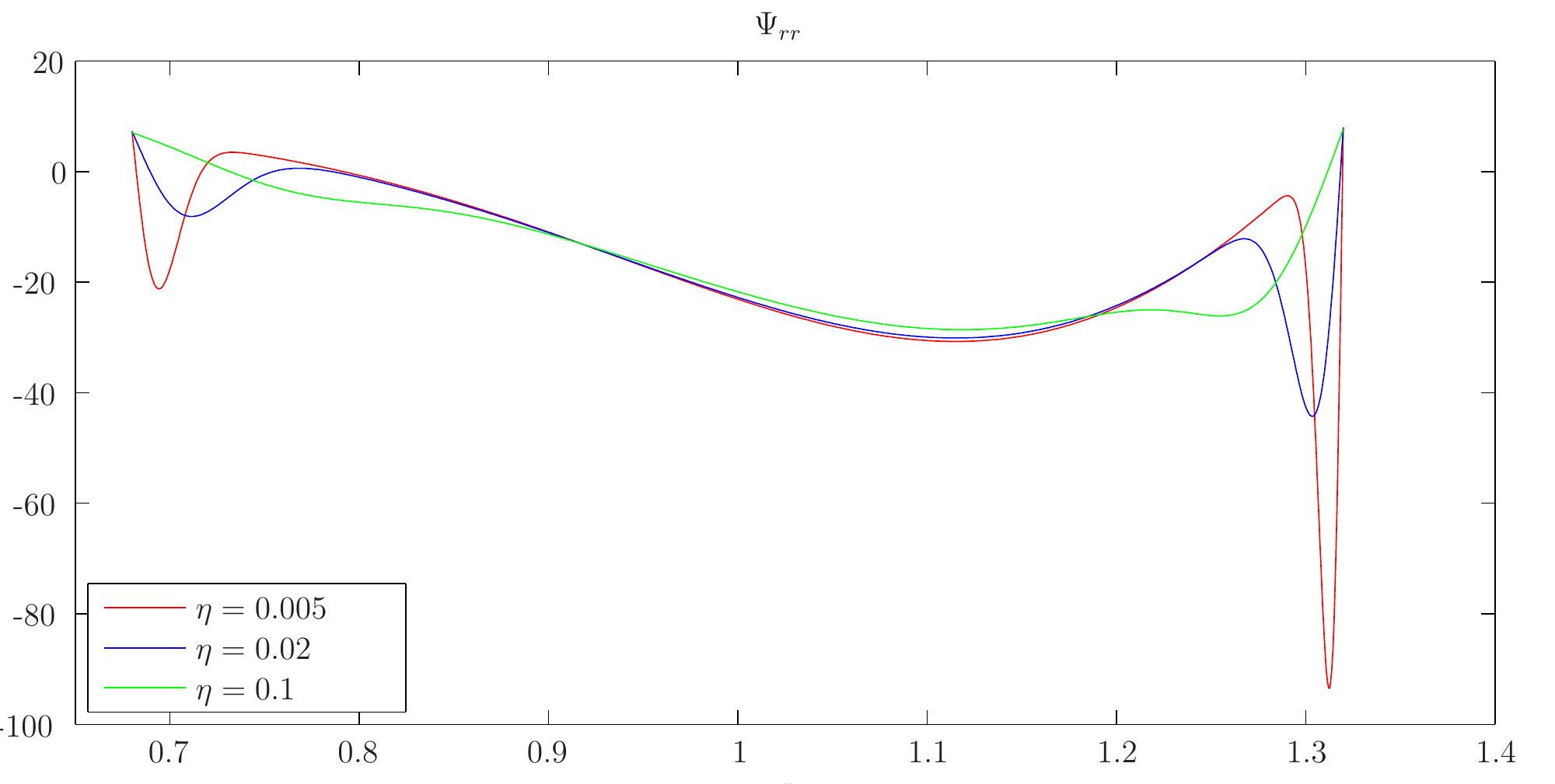}
\end{center}
\vspace{-6pt}
\caption{$\Psi, \Psi_r, \Psi_{rr} $ along $z=0$.}
\label{fig-GS-Pedestal-3}
\end{figure}

\begin{equation}\label{eq:GS_eigen_ex}
\lapl^*\Psi=\bar{\sigma} R^{2}
\left[2C_{2}\Psi\left(1-e^{-\Psi^{2}/\eta}\right)+\frac{2}{\eta}
(C_{1}+C_{2}\Psi^{2})\Psi e^{-\Psi^{2}/\eta}\right].
\end{equation}
We solve (\ref{eq:GS_eigen_ex}) for the eigenfunction $\Psi$ and 
the eigenvalue $\bar{\sigma}$
using the iterative algorithm presented in Section~\ref{sec:iteration}.
The contours of the solution for the case $\eta=0.005$ are shown in 
Figure~\ref{fig-GS-Pedestal-2}.

In order to understand the influence of the steepness of the pressure 
pedestal on the solution,
$\Psi$ and its first and second radial derivatives are plotted 
along the line $z=0$.
The results are displayed in Figure~\ref{fig-GS-Pedestal-3}.
We see that the solution $\Psi$ and its first derivative $\Psi_{R}$ 
depend only weakly on $\eta$.
However, in the vicinity of the edge of the plasma, in the pedestal region,
the behavior of the second derivative $\Psi_{RR}$ is very sensitive to the 
value of $\eta$.

\section{Conclusion}
This article describes a new, fixed-boundary, direct Grad-Shafranov solver
that relies upon conformal mapping and Green's function methods
to compute high-order accurate plasma equilibria.
Its attractive features are its proven spectral convergence, 
its speed, and its versatility in that arbitrary plasma boundaries 
can be given as input (both up-down symmetric and asymmetric). 
The solver is shown to have very good performance for a wide range of 
pressure profiles, including fusion-relevant profiles with a steep pressure 
pedestal. Equilibria with large Shafranov shifts are also computed 
without difficulty.

The high accuracy achieved for the first and second derivatives of the 
solution suggests that this solver could succesfully be combined with 
macroscopic stability codes, or implemented in transport studies on 
long time scales in which the equilibrium profiles are evolved 
self-consistently according to the GS equation. This is particularly 
true for low aspect ratio devices such as the tokamak, for which crowding 
in the conformal mapping is not very pronounced, and relatively coarse 
grids yield very accurate solutions. For high aspect ratio devices 
such as the ST, the high crowding factor observed in the conformal 
mapping suggests that alternative methods may be more efficient. 
In this regime, a Poisson solver relying on fast multipole methods
\cite{mckenney,ethridge,greengard,langston} 
or compression \cite{martinsson}, both of which
are highly adaptive, may represent an attractive alternative.
This would also permit calculations on non-smooth plasma boundaries 
with fusion-relevant $X$-points to be performed.
These alternative approaches are subjects of ongoing research.

\bibliography{preprint}

\begin{thebibliography}{44}
\providecommand{\natexlab}[1]{#1}
\providecommand{\url}[1]{\texttt{#1}}
\expandafter\ifx\csname urlstyle\endcsname\relax
  \providecommand{\doi}[1]{doi: #1}\else
  \providecommand{\doi}{doi: \begingroup \urlstyle{rm}\Url}\fi

\bibitem[Aymar et~al.(1996)Aymar, Chuyanov, Huguet, Parker, and
  Shimomura]{ITER2}
R.~Aymar, V.~Chuyanov, M.~Huguet, R.~Parker, and Y.~Shimomura.
\newblock {The ITER Project: A Physics and Technology Experiment}.
\newblock In \emph{{Proceedings of the 16th International Conference on Fusion
  Energy}}, volume~1, page~3, Montreal, Canada, 1996.

\bibitem[Aymar et~al.(2002)Aymar, Barabaschi, and Shinomura]{ITER1}
R.~Aymar, P.~Barabaschi, and Y.~Shinomura.
\newblock {The ITER design}.
\newblock \emph{Plasma Phys. Controlled Fusion}, 44\penalty0 (5):\penalty0
  519--565, 2002.

\bibitem[Berrut and Trefethen(2004)]{berrut}
J.-P. Berrut and L.~N. Trefethen.
\newblock {Barycentric Lagrange Interpolation}.
\newblock \emph{SIAM Rev.}, 46\penalty0 (3):\penalty0 501--517, 2004.

\bibitem[Beurskens et~al.(2011)Beurskens, Osborne, Schneider, Wolfrum,
  Frassinetti, Groebner, Lomas, Nunes, Saarelma, Scannell, Snyder, Zarzoso,
  Balboa, Bray, Brix, Flanagan, Giroud, Giovannozzi, Kempenaars, Loarte, {de la
  Luna}, Maddison, Maggi, McDonald, Pasqualotto, Saibene, Sartori, Solano,
  Walsh, and Zabeo]{Beurskens_ped}
M.~N.~A. Beurskens, T.~H. Osborne, P.~A. Schneider, E.~Wolfrum, L.~Frassinetti,
  R.~Groebner, P.~Lomas, I.~Nunes, S.~Saarelma, R.~Scannell, P.~B. Snyder,
  D.~Zarzoso, I.~Balboa, B.~Bray, M.~Brix, J.~Flanagan, C.~Giroud,
  E.~Giovannozzi, M.~Kempenaars, A.~Loarte, E.~{de la Luna}, G.~Maddison, C.~F.
  Maggi, D.~McDonald, R.~Pasqualotto, G.~Saibene, R.~Sartori, E.~Solano,
  M.~Walsh, and L.~Zabeo.
\newblock {H-mode pedestal scaling in DIII-D, ASDEX Upgrade, and JET}.
\newblock \emph{Phys. Plasmas}, 18\penalty0 (5):\penalty0 056120, 2011.

\bibitem[Cerfon and Freidberg(2010)]{Cerfon}
A.~J. Cerfon and J.~P. Freidberg.
\newblock {"One size fits all" analytic solutions to the Grad-Shafranov
  equation}.
\newblock \emph{Phys. Plasmas}, 17\penalty0 (3):\penalty0 032502, 2010.

\bibitem[Chen et~al.(2000)Chen, Su, and Shizgal]{chen}
H.~Chen, Y.~Su, and B.~Shizgal.
\newblock {A Direct Spectral Collocation Poisson Solver in Polar and
  Cylindrical Coordinates}.
\newblock \emph{J. Comput. Phys.}, 160\penalty0 (2):\penalty0 453--469, 2000.

\bibitem[DeLucia et~al.(1980)DeLucia, Jardin, and Todd]{DeLucia_inverse}
J.~DeLucia, S.~C. Jardin, and A.~M.~M. Todd.
\newblock {An iterative metric method for solving the inverse tokamak
  equilibrium problem}.
\newblock \emph{J. Comput. Phys.}, 37\penalty0 (2):\penalty0 183--204, 1980.

\bibitem[Ethridge and Greengard(2001)]{ethridge}
F.~Ethridge and L.~Greengard.
\newblock {A New Fast-Multipole Accelerated Poisson Solver in Two Dimensions}.
\newblock \emph{SIAM J. Sci. Comput.}, 23\penalty0 (3):\penalty0 741--760,
  2001.

\bibitem[Freidberg(1985)]{Freidberg_book}
J.~P. Freidberg.
\newblock \emph{{Ideal Magnetohydrodynamics}}.
\newblock Springer, New York, 1 edition, 1985.

\bibitem[Goedbloed(1981)]{Goedbloed_conform_1}
J.~P. Goedbloed.
\newblock {Conformal mapping methods in two-dimensional magnetohydrodynamics}.
\newblock \emph{Comput. Phys. Commun.}, 24\penalty0 (3--4):\penalty0 311--321,
  1981.

\bibitem[Goedbloed(1984)]{Goedbloed_conform_2}
J.~P. Goedbloed.
\newblock {Some remarks on computing axisymmetric equilibria}.
\newblock \emph{Comput. Phys. Commun.}, 31\penalty0 (2--3):\penalty0 123--135,
  1984.

\bibitem[Goedbloed et~al.(2010)Goedbloed, Keppens, and
  Poedts]{Goedbloed_textbook}
J.~P. Goedbloed, R.~Keppens, and S.~Poedts.
\newblock \emph{{Advanced Magnetohydrodynamics: With Applications to Laboratory
  and Astrophysical Plasmas}}.
\newblock Cambridge University Press, Cambridge, 2010.

\bibitem[Gourdain et~al.(2006)Gourdain, Leboeuf, and Neches]{Gourdain_Euler}
P.-A. Gourdain, J.-N. Leboeuf, and R.~Y. Neches.
\newblock {High-resolution magnetohydrodynamic equilibrium code for unity beta
  plasmas}.
\newblock \emph{J.~Comput. Phys.}, 216\penalty0 (1):\penalty0 275--299, 2006.

\bibitem[Grad and Rubin(1958)]{Grad}
H.~Grad and H.~Rubin.
\newblock {Hydromagnetic Equilibria and Force-Free Fields}.
\newblock \emph{United Nations Conference on the Peaceful Uses of Atomic
  Energy}, 31:\penalty0 190--197, 1958.

\bibitem[Greengard and Lee(1996)]{greengard}
L.~Greengard and J.-Y. Lee.
\newblock {A Direct Adaptive Poisson Solver of Arbitrary Order Accuracy}.
\newblock \emph{J. Comput. Phys.}, 125\penalty0 (2):\penalty0 415--424, 1996.

\bibitem[Gruber et~al.(1987)Gruber, Iacono, and Troyon]{Gruber_inverse}
R.~Gruber, R.~Iacono, and F.~Troyon.
\newblock {Computation of MHD equilibria by a quasi-inverse finite hybrid
  element approach}.
\newblock \emph{J. Comput. Phys.}, 73\penalty0 (1):\penalty0 168--182, 1987.

\bibitem[Holmes et~al.(1980)Holmes, Peng, and Lynch]{Holmes_Euler}
J.~A. Holmes, Y.-K.~M. Peng, and S.~J. Lynch.
\newblock {Evolution of Flux-Conserving Tokamak Equilibria with Preprogrammed
  Cross Sections}.
\newblock \emph{J.~Comput. Phys.}, 36\penalty0 (1):\penalty0 35--54, 1980.

\bibitem[Howell and Sovinec(2008)]{Howell_NIMEQ}
E.~C. Howell and C.~R. Sovinec.
\newblock {NIMEQ: MHD Equilibrium Solver for NIMROD}.
\newblock In \emph{{APS Meeting Abstracts}}, page 6041P, Nov. 2008.

\bibitem[Huysmans et~al.(1991)Huysmans, Goedbloed, and Kerner]{Huysmans_Euler}
G.~T.~A. Huysmans, J.~P. Goedbloed, and W.~Kerner.
\newblock {Isoparametric bicubic Hermite elements for solution of the
  Grad–Shafranov equation}.
\newblock \emph{Proc. CP90 Conf. on Comp. Phys. Proc.}, pages 371--376, 1991.

\bibitem[Jardin(2010)]{Jardin_textbook}
S.~Jardin.
\newblock \emph{{Computational Methods in Plasma Physics}}.
\newblock Chapman \& Hall / CRC Press, New York, 1 edition, 2010.

\bibitem[Jardin(2004)]{Jardin_Euler}
S.~C. Jardin.
\newblock {A triangular finite element with first-derivative continuity applied
  to fusion MHD applications}.
\newblock \emph{J.~Comput. Phys.}, 200\penalty0 (1):\penalty0 133--152, 2004.

\bibitem[Kerzman and Stein(1978)]{Kerzman_Stein}
N.~Kerzman and E.~M. Stein.
\newblock {The Cauchy kernel, the Szeg\"o kernel, and the Riemann mapping
  function}.
\newblock \emph{Mathematische Annalen}, 236\penalty0 (1):\penalty0 85--93,
  1978.

\bibitem[Kerzman and Trummer(1986)]{Kerzman_Trummer}
N.~Kerzman and M.~R. Trummer.
\newblock {Numerical conformal mapping via the Szeg\"o kernel}.
\newblock \emph{J. Comput. Appl. Math.}, 14\penalty0 (1--2):\penalty0 111--123,
  1986.

\bibitem[Langston et~al.(2011)Langston, Greengard, and Zorin]{langston}
H.~Langston, L.~Greengard, and D.~Zorin.
\newblock {A Free-Space Adaptive FMM-Based PDE Solver in Three Dimensions}.
\newblock \emph{Comm. Appl. Math. and Comp. Sci.}, 6\penalty0 (1):\penalty0
  79--122, 2011.

\bibitem[Ling and Jardin(1985)]{Ling_inverse}
K.~M. Ling and S.~C. Jardin.
\newblock {The Princeton Spectral Equilibrium Code: PSEC}.
\newblock \emph{J. Comput. Phys.}, 58\penalty0 (3):\penalty0 300--335, 1985.

\bibitem[Lo{D}estro and Pearlstein(1994)]{Lodestro}
L.~L. Lo{D}estro and L.~D. Pearlstein.
\newblock {On the Grad-Shafranov equation as an eigenvalue problem, with
  implications for q solvers}.
\newblock \emph{Phys. Plasmas}, 1\penalty0 (1):\penalty0 90--95, 1994.

\bibitem[Ludwig(1997)]{Ludwig_inverse}
G.~O. Ludwig.
\newblock {Direct variational solutions of the tokamak equilibrium problem}.
\newblock \emph{Plasma Phys. Controlled Fusion}, 39:\penalty0 2021--2037, 1997.

\bibitem[L\"ust and Schl\"uter(1957)]{Schluter}
R.~L\"ust and A.~Schl\"uter.
\newblock {Axial symmetrische magnetohydrodynamische
  Gleichgewichtskonfigurationen}.
\newblock \emph{Z.~Naturforsch.}, 12a:\penalty0 850--854, 1957.

\bibitem[L\"utjens et~al.(1992)L\"utjens, Bondeson, and Roy]{Lutjens_Euler}
H.~L\"utjens, A.~Bondeson, and A.~Roy.
\newblock {Axisymmetric MHD equilibrium solver with bicubic Hermite elements}.
\newblock \emph{Comput. Phys. Commun.}, 69\penalty0 (2--3):\penalty0 287--298,
  1992.

\bibitem[L\"utjens et~al.(1996)L\"utjens, Bondeson, and
  Sauter]{Lutjens_CHEASE2}
H.~L\"utjens, A.~Bondeson, and O.~Sauter.
\newblock {The CHEASE code for toroidal MHD equilibria}.
\newblock \emph{Comput. Phys. Commun.}, 97\penalty0 (3):\penalty0 219--260,
  1996.

\bibitem[Maddison et~al.(2009)Maddison, Hubbard, Hughes, Snipes, LaBombard,
  Nunes, Beurskens, Erents, Kempenaars, Alper, Pinches, Valovi{\v c},
  Pasqualotto, Alfier, and Giovannozzi]{Maddison_ped}
G.~Maddison, A.~Hubbard, J.~Hughes, J.~Snipes, B.~LaBombard, I.~Nunes,
  M.~Beurskens, S.~Erents, M.~Kempenaars, B.~Alper, S.~Pinches, M.~Valovi{\v
  c}, R.~Pasqualotto, A.~Alfier, and E.~Giovannozzi.
\newblock {Dimensionless pedestal identity plasmas on Alcator C-Mod and Jet}.
\newblock \emph{Nucl. Fusion}, 49\penalty0 (12):\penalty0 125004, 2009.

\bibitem[Martinsson(2009)]{martinsson}
P.-G. Martinsson.
\newblock {A fast direct solver for a class of elliptic partial differential
  equations}.
\newblock \emph{J. Sci. Comput.}, 38\penalty0 (3):\penalty0 316--330, 2009.

\bibitem[McKenney et~al.(1995)McKenney, Greengard, and Mayo]{mckenney}
A.~McKenney, L.~Greengard, and A.~Mayo.
\newblock {A fast Poisson solver for complex geometries}.
\newblock \emph{J. Comput. Phys.}, 118\penalty0 (2):\penalty0 348--355, 1995.

\bibitem[Odonnell and Rokhlin(1989)]{Odonnell}
S.~T. Odonnell and V.~Rokhlin.
\newblock {A fast algorithm for the numerical evaluation of conformal
  mappings}.
\newblock \emph{SIAM J. Sci. Stat. Comput.}, 10\penalty0 (3):\penalty0
  475--487, 1989.

\bibitem[Pataki(2011)]{Pataki}
A.~Pataki.
\newblock \emph{{High-order methods for elliptic problems in plasma physics}}.
\newblock PhD thesis, Courant Institute of Mathematical Sciences, New York
  University, 2011.

\bibitem[Sabbagh et~al.(2001)Sabbagh, Kaye, Menard, Paoletti, Bell, Bell,
  Bialek, Bitter, Fredrickson, Gates, Glasser, Kugel, Lao, LeBlanc, Maingi,
  Maqueda, Mazzucato, Mueller, Ono, Paul, Peng, Skinner, Stutman, Wurden, Zhu,
  and {NSTX Research team}]{NSTX}
S.~A. Sabbagh, S.~M. Kaye, J.~Menard, F.~Paoletti, M.~Bell, R.~E. Bell,
  J.~Bialek, M.~Bitter, E.~D. Fredrickson, D.~A. Gates, A.~H. Glasser,
  H.~Kugel, L.~L. Lao, B.~P. LeBlanc, R.~Maingi, R.~J. Maqueda, E.~Mazzucato,
  D.~Mueller, M.~Ono, S.~F. Paul, M.~Peng, C.~H. Skinner, D.~Stutman, G.~A.
  Wurden, W.~Zhu, and {NSTX Research team}.
\newblock {Equilibrium properties of spherical torus plasmas in NSTX}.
\newblock \emph{Nucl. Fusion}, 41\penalty0 (11):\penalty0 1601--1611, 2001.

\bibitem[Shafranov(1958)]{Shafranov}
V.~D. Shafranov.
\newblock On magnetohydrodynamical cquilibrium configurations.
\newblock \emph{Sov. Phys. JETP}, 6\penalty0 (3):\penalty0 545--554, 1958.

\bibitem[Shen(1997)]{shen}
J.~Shen.
\newblock {Efficient Spectral-Galerkin Methods III: Polar and Cylindrical
  Geometries}.
\newblock \emph{SIAM J. Sci. Comput.}, 18\penalty0 (6):\penalty0 1583--1604,
  1997.

\bibitem[Solov'ev(1968)]{Solovev}
L.~S. Solov'ev.
\newblock {The Theory of Hydromagnetic Stability of Toroidal Plasma
  Configurations}.
\newblock \emph{Sov. Phys. JETP}, 26\penalty0 (2):\penalty0 400--407, 1968.

\bibitem[Takeda and Tokuda(1991)]{Takeda_review}
T.~Takeda and S.~Tokuda.
\newblock {Computation of MHD equilibrium of tokamak plasma}.
\newblock \emph{J.~Comput. Phys.}, 93\penalty0 (1):\penalty0 1--107, 1991.

\bibitem[Trefethen and Bau(1997)]{Trefethen}
L.~Trefethen and D.~Bau.
\newblock \emph{{Numerical linear algebra}}.
\newblock SIAM, 1997.

\bibitem[Turkington et~al.(1993)Turkington, Lifschitz, Eydeland, and
  Spruck]{Turkington_inverse}
B.~Turkington, A.~Lifschitz, A.~Eydeland, and J.~Spruck.
\newblock {Multiconstrained variational problems in magnetohydrodynamics:
  Equilibrium and slow evolution}.
\newblock \emph{J. Comput. Phys.}, 106\penalty0 (2):\penalty0 269--285, 1993.

\bibitem[Zakharov and Pletzer(1999)]{Zakharov}
L.~Zakharov and A.~Pletzer.
\newblock {Theory of perturbed equilibria for solving the Grad-Shafranov
  equation}.
\newblock \emph{Phys. Plasmas}, 6\penalty0 (12):\penalty0 4693--4704, 1999.

\bibitem[Zheng et~al.(1996)Zheng, Wootton, and Solano]{Zheng}
S.~B. Zheng, A.~J. Wootton, and E.~R. Solano.
\newblock {Analytical tokamak equilibrium for shaped plasmas}.
\newblock \emph{Phys. Plasmas}, 3\penalty0 (3):\penalty0 1176--1178, 1996.

\end{thebibliography}
\bibliographystyle{abbrvnat}

\end{document}